\documentclass[%
 aip,
 amsmath,amssymb,
 reprint,%
]{revtex4-1}

\usepackage{graphicx}
\usepackage{dcolumn}
\usepackage{bm}

\usepackage[utf8]{inputenc}
\usepackage[T1]{fontenc}
\usepackage{mathptmx}
\usepackage{etoolbox}

\usepackage{hyperref}
\usepackage{physics}

\makeatletter
\def\@email#1#2{%
 \endgroup
 \patchcmd{\titleblock@produce}
  {\frontmatter@RRAPformat}
  {\frontmatter@RRAPformat{\produce@RRAP{*#1\href{mailto:#2}{#2}}}\frontmatter@RRAPformat}
  {}{}
}%
\makeatother

\begin{document}

\title{Quantum chaos on the separatrix of the periodically perturbed Harper model}
\author{Alice C. Quillen}
\author{Abobakar Sediq Miakhel} 
\affiliation{$^1$Department of Physics and Astronomy, University of Rochester,  Rochester, NY, 14627, USA}
\email{aquillen@ur.rochester.edu,amiakhel@u.rochester.edu}
\date{\today}

\begin{abstract}
We explore the relation between a classical periodic Hamiltonian system and an associated discrete quantum system on a torus in phase space.  The model is a sinusoidally perturbed Harper model and is similar to the sinusoidally perturbed pendulum.   Separatrices connecting hyperbolic fixed points in the unperturbed classical system become chaotic under sinusoidal perturbation.  We numerically compute eigenstates of the Floquet propagator for the associated quantum system.  For each propagator eigenstate we compute a Husimi distribution in phase space and an energy and energy dispersion from the expectation value of the unperturbed Hamiltonian operator.  The Husimi distribution of each Floquet eigenstate resembles a classical orbit with a similar energy and similar energy dispersion. Chaotic orbits in the mixed classical system are related to Floquet eigenstates that appear ergodic. For a mixed regular and chaotic system,  the energy dispersion can separate the Floquet eigenstates into ergodic and integrable subspaces.  The width of a chaotic region in the classical system is estimated by integrating the perturbation along a separatrix orbit. We derive a related expression for the associated quantum system from the averaged perturbation in the interaction representation evaluated at states with energy close to the separatrix. 



\end{abstract}

\maketitle

\begin{quotation}


We explore the relation between closely related periodically varying classical and quantum systems containing both regular and chaotic behavior.  
Our systems are confined in phase space, allowing a finite dimension for the associated quantum system,    facilitating computation of the Floquet propagator, and giving potential applications in quantum computing. 
Husimi distributions showing the structure of eigenstates in phase space of the quantum system are remarkably similar to classical orbits.  We find that the dispersion of the unperturbed energy function or operator locates ergodicity in both classical and quantum settings. 
We estimate the width of the ergodic region in the quantum system, deriving a quantum counterpart 
to a classical estimate. 
\end{quotation}


\section{Introduction \label{sec:intro}}

Classical Hamiltonian dynamical systems can exhibit remarkable complexity
 with both regular and chaotic dynamics (e.g., \citet{Chirikov_1979}).  
A quantum mechanical system, that is derived from or related to a classical one, can 
 exhibit phenomena that is related to the classical dynamics. 
Eigenfunctions, called quantum scars, can be found in the vicinity of the periodic orbits
of the classical system \citep{Heller_1984,Berry_1989} and along  
a separatrix orbit \citep{Scharf_1992}. 
It has been conjectured \citep{Bohigas_1984} that the eigenvalue statistics of a quantum system derived from a classical chaotic one can be described by random matrix theory, whereas the eigenvalue statistics of a quantum system 
derived from a classical integrable system follows Poisson statistics \citep{Berry_1977}. 
At large kick strengths, 
the quantum kicked rotor, related to the classical kicked rotor and the standard map,  can 
be mapped to an Anderson-type model with a quasi-disordered
potential \citep{Fishman_1982} and exhibits a quasi-energy spectrum resembling the Wigner-Dyson form that is characteristic of the Gaussian orthogonal ensemble (GOE) random matrix model  \citep{Izrailev_1986,Chirikov_1988}. 

Time-periodic driving is a potentially powerful tool for controlling the material properties and dynamics 
of quantum systems  \citep{Floquet_2020} and for carrying out quantum computations \citep{Farhi_2000,Albash_2018}.  
Chaos-­assisted quantum tunneling is a process where coupling between quantum states located in
 regular islands in phase space is mediated by ergodic states that are 
spread out over a chaotic region \citep{Arnal_2020,Hanada_2023}. 
In this paper we are interested in the relationship between periodic chaotic classical systems and 
their quantum discrete counterparts.  We choose to study evolution in finite dimensional quantum systems 
to facilitate numerical computations and because these systems would be relevant for control of 
quantum systems used for quantum computing. 

The Melnikov method establishes the existence of chaotic behavior in a classical Hamiltonian dynamical system under periodic perturbation  \citep{Melnikov_1963}.   The integral of a sinusoidal 
 perturbation near the separatrix orbit of the unperturbed system can be used to generate a map, 
 known as a separatrix map, giving an estimate of the width of a chaotic region 
\citep{Zaslavsky_1968,Chirikov_1979,Shevchenko_2000,Soskin_2008,Soskin_2009,Treschev_2010}.   
This approach was used 
to derive a condition for chaotic behavior known as the Chirikov resonance-overlap criterion \citep{Chirikov_1979}.  Quantized versions of separatrix maps  
can exhibit dynamical localization \citep{Bubner_1991,Iomin_2003}.
\citet{Yampolsky_2022} in their study of a perturbed quantum system of hard core particles in a box,  recently have shown that 
the presence of quantum chaos can be estimated from a classical resonance-overlap criterion. 
We similarly explore the possibility that classical techniques for estimating the location of chaotic regions in 
phase have quantum counterparts, but in periodically perturbed systems which are also known as 
Floquet systems (e.g., \citet{Neufeld_2019}). 

We desire a finite dimensional Hamiltonian model that that we can visualize and quantize,  and that exhibits adjustable regular and chaotic behavior.  
We choose a variant of the periodically perturbed pendulum model 
that was used by \citet{Chirikov_1979} to derive the  
 criterion, known as the resonance overlap criterion,
 for the onset of chaotic motion in deterministic classical Hamiltonian systems. 
 That particular system is a pendulum, with Hamiltonian $H_0(\phi,p) = a\frac{p^2}{2} - \epsilon \cos \phi $, 
 that is perturbed with a periodic function, giving 
the time dependent and periodic Hamiltonian  (equation 4.50 by \citet{Chirikov_1979});  
 \begin{align}
H(\phi,p,t )_{\rm PertPend} & =a\frac{p^2}{2} - \epsilon \cos \phi  - \mu ( \cos (\phi + \nu t)\nonumber \\
& \qquad  + \cos (\phi - \nu t)). \label{eqn:PertPend}
\end{align} 
A weak perturbation, with $\mu < \epsilon$,  drives instability 
at the separatrix of the unperturbed system.   The system has 
 three resonances, centered at $p =0, \pm \nu/a$  with strengths set by $\epsilon$ and $\mu$, respectively.
 The frequency of libration about the fixed point at $p=0, \phi=0$ in the unperturbed system (with $\mu = \mu'=0$)  is 
  \begin{align}
 \omega_0 = \sqrt{\epsilon a}.
 \end{align} 
The coordinate angle $\phi \in [0, 2 \pi)$ is periodic, but the momentum $p \in \mathbb{R}$. 
The cosine potential of the unperturbed system (the term $\propto \cos \phi$ in equation \ref{eqn:PertPend})  has been achieved with counter-propagating laser beams
on a rubidium-87 Bose-Einstein
condensate machine \citep{Arnal_2020} and is equivalent to the Josephson energy of a transmon 
(e.g., \citep{Cohen_2023}).  

The perturbed pendulum model of equation \ref{eqn:PertPend} is similar to the kicked rotor; 
 \begin{align}
H(\phi,p,t )_{\rm KickedRotor} = \frac{p^2}{2} - \epsilon \cos \phi D_T(t) \label{eqn:Harper}
\end{align}
where $D_T(t)$  is the Dirac comb, consisting of an infinite sum of delta functions, each separated by
period $T$.
The kicked rotor, when mapped at the period $T$ and with momentum $p$ restricted to a periodic interval,
generates the well known {\it standard map}, also introduced by \citet{Chirikov_1979}. 

The kicked rotor can be quantized by replacing $ \phi, p$ with linear operators that operate on 
vectors in a Hilbert space,  \citep{Casati_1979, Santhanam_1998}. 
The quantum model can be restricted to a discrete (or finite dimensional) rather than continuous Hilbert space 
by choosing eigenstate solutions that are periodic in $p$ \citep{Chirikov_1988}. 

The Harper model, with Hamiltonian $H(\phi,p) = \cos p + \cos \phi$,  
is a doubly periodic model used to describe the motion 
of electrons in a 2-dimensional lattice in the presence of a magnetic field \citep{Harper_1955}. 
While the Harper model is regular (integrable because it is 2-dimensional and the Hamiltonian is conserved),  
a generalization of this model, called the kicked
Harper model, can exhibit chaotic behavior  \citep{Lebouf_1998}; 
\begin{align}
H(\phi,p,t)_{\rm KickedHarper} = A \cos p + B \cos \phi  D_T(t) ,  \label{eqn:kicked_harper}
\end{align} 
where $D_T(t)$ is again the Dirac comb
and $A,B$ are real coefficients. 
The kicked Harper model exhibits both regular and
chaotic orbits, similar to the kicked rotor, but its phase space is a torus $\phi, p \in [0, 2 \pi)$ (e.g., \citet{Levi_2004}).    
Usually the phase space for the Harper model is a coordinate and a momentum, both
confined on a periodic lattice, but here we have written the coordinate as an angle, so the associated momentum is an action variable and the model looks similar to the kicked rotor.  
For small $p$, the kinetic term $1-\cos p \sim \frac{p^2}{2}$ and in the same form as 
 the kinetic energy of a harmonic oscillator or pendulum.  
As is true for the quantized kicked rotor, in the kicked Harper model, 
when the perturbation is high enough that the associated classical system is chaotic, 
the statistics of the quasi-energies of the Floquet propagator are similar to that
of a random matrix model \citep{Wei_1991}. 

An advantage of the Harper model is that its Hamiltonian is periodic in both momentum and coordinate, so 
phase space is compact and equivalent to a torus.   When quantized, there is a finite number of 
quantum states facilitating numerical computations.    In contrast, 
for the kicked rotor with unrestricted momentum, the system must be truncated in Fourier space to 
 compute the Floquet propagator (e.g., \citet{Izrailev_1989}).     
In our study we modify the Harper model to resemble the perturbed pendulum of equation 
\ref{eqn:PertPend}.  This gives a classical Hamiltonian model on the torus with an adjustable chaotic separatrix width that we can compare with discrete quantized versions. 

\begin{figure*}[htbp]\centering
\includegraphics[width=2.1 truein, trim=15 15 10 10,clip]{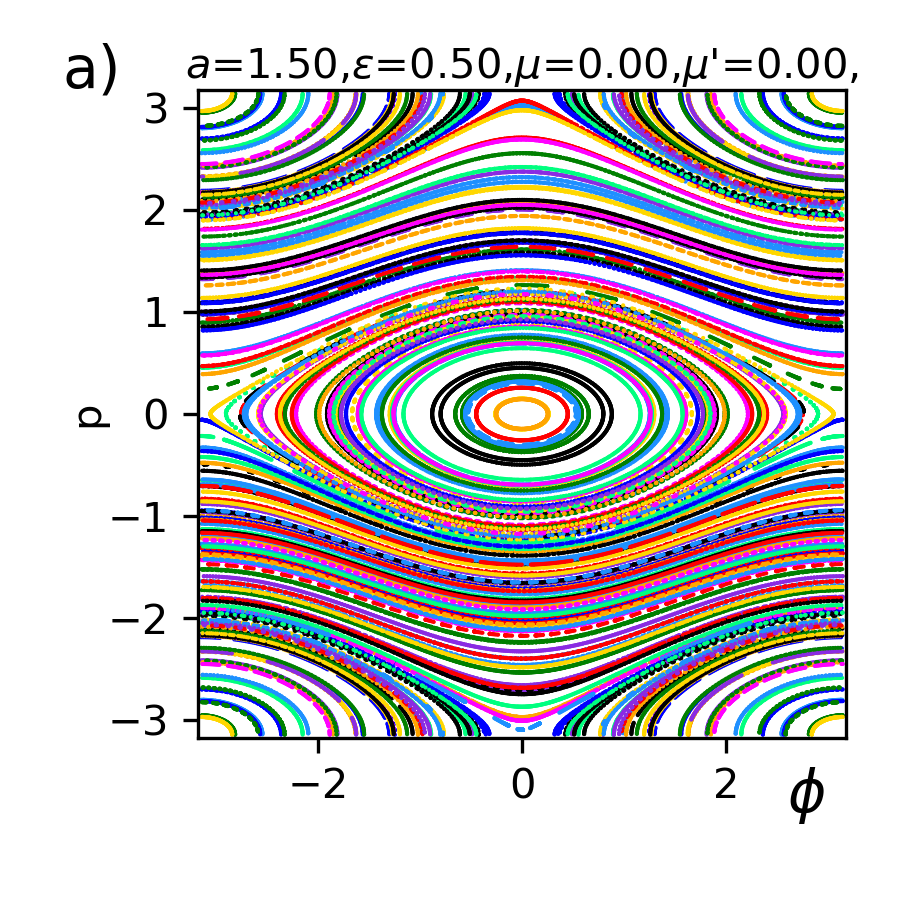}
\includegraphics[width=2.1 truein,trim = 0 -10 0 0,clip]{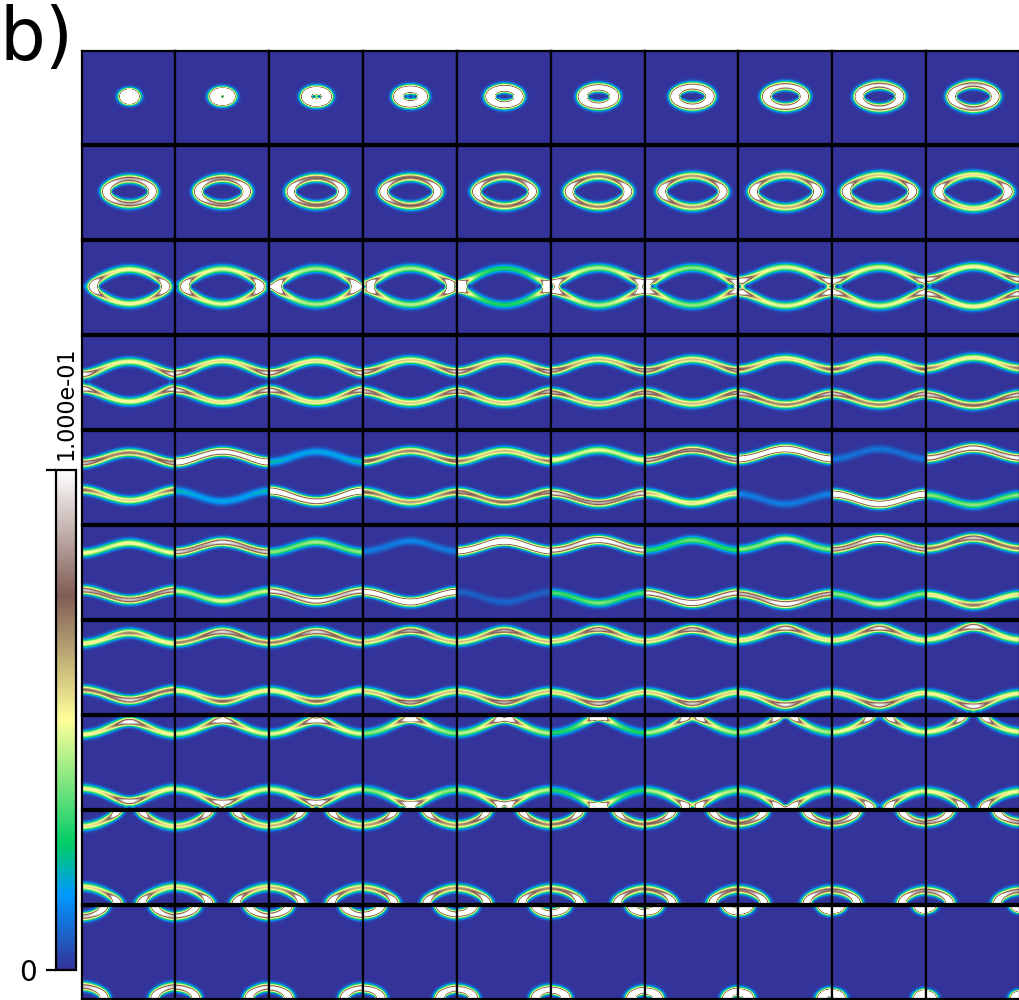}
\includegraphics[width=2.3 truein,trim = 0 0 0 0,clip]{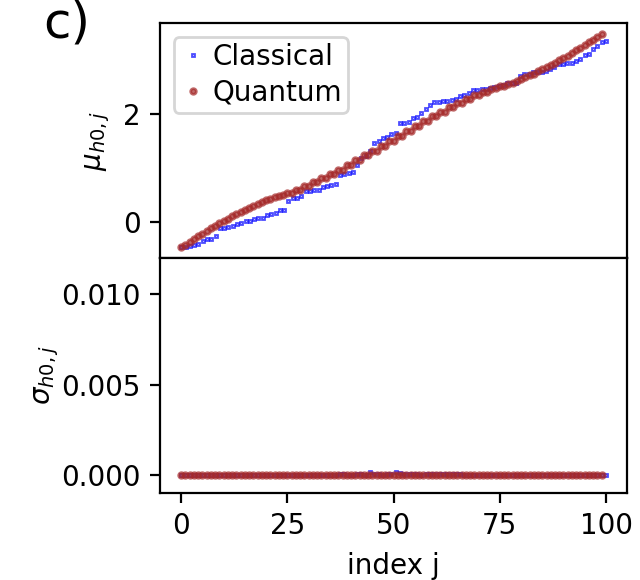}
\caption{a) We show a surface of section for regular (not chaotic) classical system, 
the Harper model with Hamiltonian in equation \ref{eqn:Hclassical}), 
and with parameters of the model printed on top of the figure.   
Points are plotted once every perturbation period of $2\pi$.  Each orbit is 
plotted with the same color points, and we choose different colors for different orbits. 
This particular system has $\mu  = \mu' = 0$ and lacks sinusoidal time-dependent perturbations, 
but the figures are made in the same way as subsequent figures that have periodic perturbations. 
b) The related quantum model (with the same parameters $a, \epsilon, \mu, \mu'$ as the classical one) 
has Hamiltonian operator in equation \ref{eqn:Hquantum} and  
Floquet propagator given in equation \ref{eqn:U_T}. We show 
Husimi distributions for
the eigenstates of the unitary propagator for the discrete quantized model with $N=100$.  
Eigenstates of the propagator are arranged  in order of their expectation value 
$\mu_{h0,j} = \langle \hat h_0 \rangle $.    
c) The top panel shows with solid brown dots the expectation value of the unperturbed Hamiltonian $\hat h_0$  for each eigenstate of the Floquet propagator.  
The mean values of the unperturbed Hamiltonian  $\mu_{h0,j} = \langle H_0 \rangle$ 
are also computed for classically integrated orbits and these are shown as blue squares. 
The x-axis shows the state index $j$ denoted a state for the quantum system, but the classical 
values are plotted over the same range in order of increasing mean value $\langle H_0 \rangle$, 
averaged from points in an orbit.
The bottom panel shows (with brown dots)  the standard deviations $\sigma_{h0,j} $ (equation \ref{eqn:sigh0})
of the eigenvectors of the propagator.  The standard deviation of the unperturbed Hamiltonian function 
 for the same classical orbits as plotted in the top panel are plotted as blue squares. 
 \label{fig:vanil} }
\end{figure*}

\subsection{Classical Sinusoidally Perturbed Harper Model}

We take a base Harper model (equation \ref{eqn:Harper}) but add to it 
perturbations of the perturbed pendulum model (equation \ref{eqn:PertPend}).
We restore units so that we can later compare classical
and quantum system with evolution that depends upon the reduced Planck's constant, $\hbar$. 
The perturbed Harper model, 
\begin{align}
 K(\phi,L, t)_{\rm PH} &= a_K \left(1- \cos\left( \frac{L}{L_0} \right) \right) - \epsilon_K \cos(\phi) \nonumber \\
                    &   \qquad  - \mu_K \cos(\phi - \nu t) - \mu_K'\cos(\phi + \nu t), 
 \label{eqn:Kclassical}
 \end{align}    
 where $L_0$ is a scale for the momentum and 
  the real coefficients $a_K, \epsilon_K, \mu_K, \mu_K'$ have units of energy.   The Hamiltonian is periodic in time with period $T = 2\pi/\nu$.  If we choose a moment of inertia $I$ that gives 
 $a_K = L_0^2/I$, the Hamiltonian is similar to a rotor and 
with small $L$, the term $ a_K \left(1- \cos\left( \frac{L}{L_0} \right) \right) \approx  \frac{L^2}{2I}$ 
is equal to the rotor's rotational kinetic energy. 

We define a dimensionless momentum $p =L/L_0$ and adopt a dimensionless variable for time $\tau = \nu t$.  
Henceforth we work with time in units of the inverse of the perturbation frequency $\nu$. 
The Hamiltonian for the perturbed Harper model 
\begin{align}
 H(\phi,p, \tau)_{\rm PH} &= H_0(\phi,p) + H_1 (\phi,\tau) \nonumber \\
H_0(\phi,p) & =  a \left(1- \cos p \right) - \epsilon \cos(\phi) \nonumber \\
H_1(\phi,\tau) &  =  - \mu \cos(\phi - \tau) - \mu' \cos(\phi + \tau) . 
 \label{eqn:Hclassical}
 \end{align}   
The equations of motion given by Hamilton's equations are consistent with those
 of equation \ref{eqn:Kclassical},  with dimensionless 
parameters, 
\begin{align}
a = \frac{a_K}{L_0\nu},  \qquad 
\epsilon = \frac{\epsilon_K}{L_0\nu},  \qquad 
\mu= \frac{\mu_K}{L_0\nu},  \qquad 
\mu'= \frac{\mu_K'}{L_0\nu} . \label{eqn:dofs}
\end{align}  
Phase space now consists of the torus with  $\phi,p \in [0, 2 \pi)$. 

For the unperturbed system (with $\mu=\mu'=0$), 
the frequency of libration around the stable fixed point at $p=0, \phi=0$ is $\omega_0 = \sqrt{\epsilon a}$. 

For $a,\epsilon>0$ 
the maximum and minimum energy of the unperturbed system  are 
$ 2a + \epsilon$ and  $- \epsilon$. 
As is true for the pendulum, given an energy, the orbit  
$\phi(\tau),p(\tau)$ of the unperturbed system can be written explicitly in terms of elliptic functions, however 
we have not found concise expressions for them.    

Using Hamilton's equations,  the 
fixed points of the unperturbed Harper system given by $H_0$ (in equation \ref{eqn:Hclassical})
are at $(\phi^*,p^*) = (0,0), (0,\pi), (\pi,0), (\pi,\pi)$. 
Table \ref{tab:fixed} lists these fixed points  
their energies and whether they are stable or hyperbolic. 

\begin{table}
\caption{Fixed points of the Harper model \label{tab:fixed}}
\begin{tabular}{lll}
\hline
$(\phi^*,p^*)$ & $E$ & type \\
\hline
$(0,0)$     & $-\epsilon$     & stable \\
$(\pi,\pi)$ & $2a+ \epsilon$ & stable \\
$(\pi,0)$   & $\epsilon$       & hyperbolic \\
$(0,\pi)$   & $2a-\epsilon$   & hyperbolic \\
\hline
\end{tabular}
\end{table}

\begin{figure*}[htbp]\centering
\includegraphics[width=2.1 truein, trim=15 15 10 10,clip]{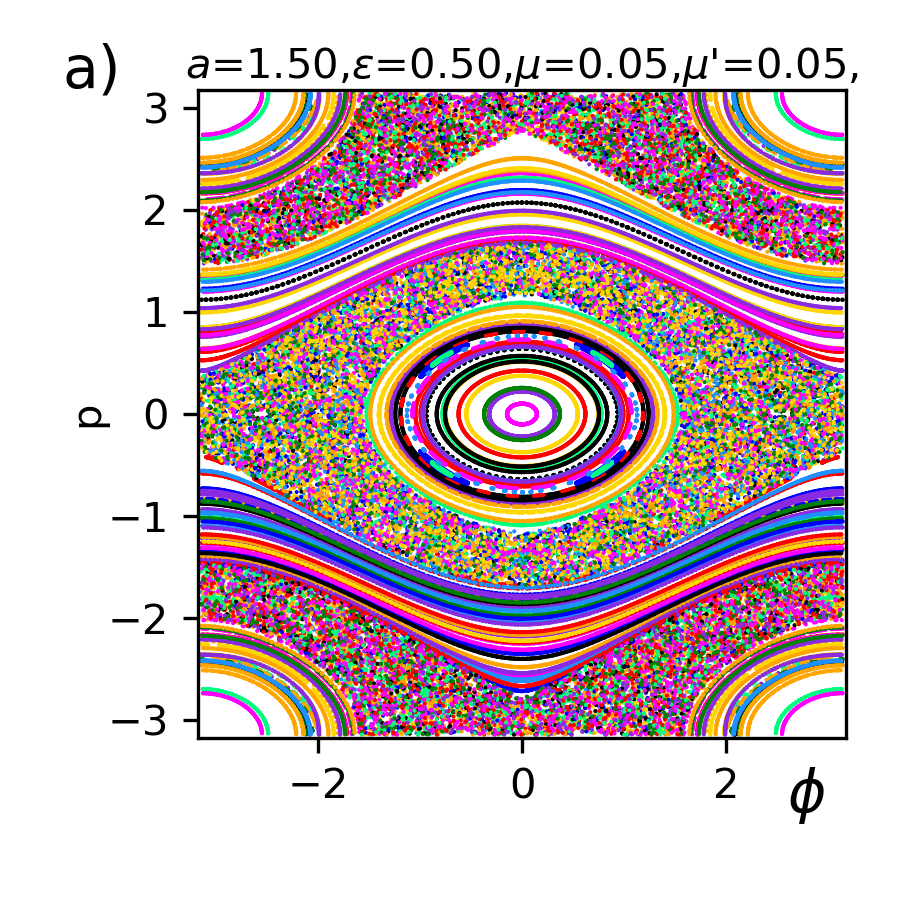}
\includegraphics[width=2.1 truein,trim = 0 -10 0 0,clip]{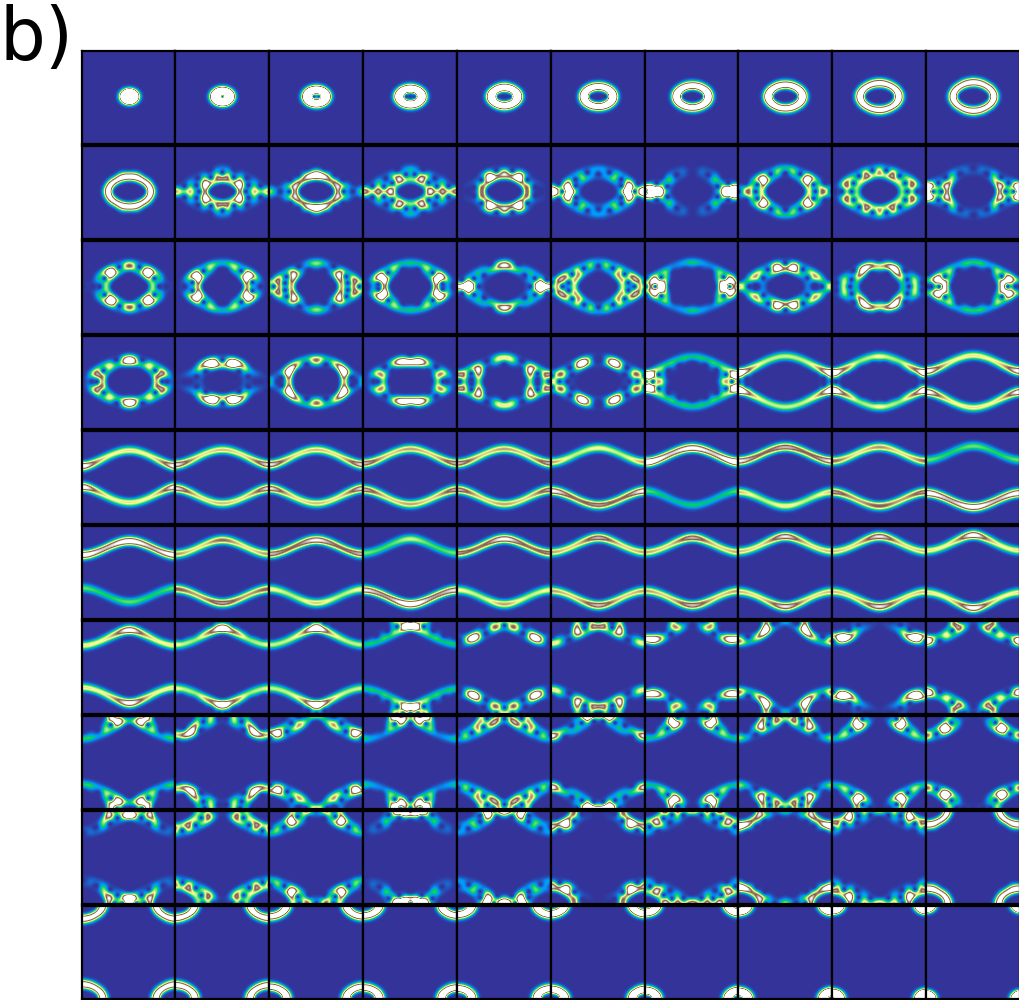}
\includegraphics[width=2.3 truein,trim = 0 0 0 2,clip]{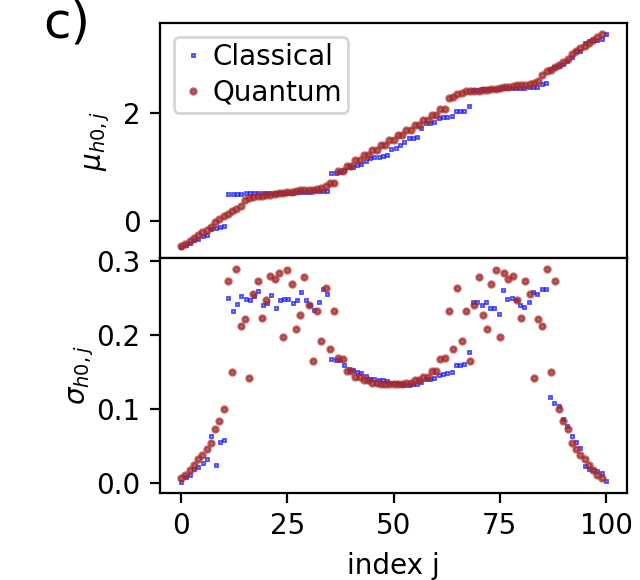}
\caption{a) Similar to Figure \ref{fig:vanil}a except for a classical 
 system that has chaotic regions.  The parameters $a,\epsilon,\mu, \mu'$ of
the Hamiltonian  are printed on the top left.  
b) We show the Hussimi distributions of the eigenstates of the associated
quantum model with $N=100$. 
The Hussimi distributions of some eigenstates of the Floquet propagator
 fill the chaotic regions in phase space seen in the associated 
classical system. The quantum system has $N=100$ states.  
c) Similar to Figure \ref{fig:vanil}c. 
We show energy means and dispersions from both classical orbits
and eigenstates. Chaotic eigenstates have large 
standard deviations $\sigma_{h0,j}$ (defined in equation \ref{eqn:sigh0}).   
Chaotic orbits also have large energy standard deviations. 
\label{fig:Q9b}}
\end{figure*}

In each of Figures \ref{fig:vanil}a, \ref{fig:Q9b}a and \ref{fig:Q9c}a, 
we show a surface of section or Poincar\'e map constructed from integrations of 200 different orbits of the Hamiltonian of equation \ref{eqn:Hclassical} (the additional panels of these figures are discussed below in section \ref{sec:num}).  
Initial conditions for the orbits are randomly chosen from uniform distributions covering phase space. 
Each orbit is plotted in one of 10 different colors and contains 500 points, with a point plotted
each period (at $\tau = 0$ modulo $2 \pi$).  In Figure \ref{fig:vanil}a,  the classical Hamiltonian 
is regular in the sense that each orbit is quasiperiodic or a curve (also called a torus when referring 
to a surface in a higher dimensional space that is filled by the trajectory).  In Figures \ref{fig:Q9b}a and \ref{fig:Q9c}a, the parameters 
$a, \epsilon, \mu, \mu'$ 
are chosen to show one or more area filling
chaotic regions that separate regions containing quasi-periodic orbits.  

When $a = \epsilon$, there is a single separatrix energy, associated with the energy at the hyperbolic fixed points (listed in Table \ref{tab:fixed})  and the separatrix forms a diamond 
along trajectories $p(\phi)_s = \pm \phi \pm \pi $.   When $a \ne \epsilon$ there are two separatrix 
energies $E_s$ associated with separate orbits  
\begin{align}
\begin{array}{llll}
\cos \frac{p_s(\phi)}{2} & =\pm  \sqrt{\frac{\epsilon}{a}} \sin \frac{\phi}{2}  &   {\rm for} &  E_s = 2a-\epsilon  \\
\sin \frac{p_s(\phi)}{2} & = \pm \sqrt{\frac{\epsilon}{a}} \cos \frac{\phi}{2}  & {\rm for} &  E_s = \epsilon .
\end{array} 
\label{eqn:sep_orbits}
\end{align} 
These orbits divide the regions of phase space in the model shown in Figure \ref{fig:vanil}a. 
The separatrix orbits are the regions that become ergodic under perturbation, as illustrated 
in Figures \ref{fig:Q9b}a and \ref{fig:Q9c}a. 

\section{A quantum mechanical version of the perturbed Harper model \label{sec:qm}}

We consider the classical Hamiltonian of equation \ref{eqn:Hclassical}.  
To facilitate calculation associated quantum  system, we choose to work in a discrete rather than 
continuous quantum space.  
We approximate the system with an $N$ dimensional quantum space 
(an $N$ dimensional normed complex vector space).  We choose a set of 
orthonormal basis 
states labelled with integers; $\ket{j}$ for $j \in \mathbb{Z}_N$ where $\mathbb{Z}_N$ is the set of 
integers $\{0, 1, ...., N-1\}$.  
The operator associated with measurement
of the angle $\phi $ we associate with this basis 
\begin{align}
\hat \phi =\sum_{j=0}^{N-1} \frac{2 \pi j }{N} \ket{j}\bra{j}   . \label{eqn:phihat}
\end{align}
Because the classical system of equation \ref{eqn:Hclassical} 
is also periodic in $p$, we simultaneously quantize the momentum $p$. 
In the limit of large $N$, we should recover the behavior of a quantum system 
in a continuous Hilbert space with angle $\phi \in [0, 2 \pi)$ and periodic. 

Two bases are mutually unbiased  
 if every measurement outcome is equally probable when a system in a state 
that is a basis element in one basis
is measured with an operator that is diagonal in the other basis (e.g., \citet{Wootters_2006}). 
In the infinite dimensional case, position and momentum operators, $\hat q, \hat p$ are associated with two bases that are mutually unbiased and related by Fourier transform.  
Given a particular orthonormal basis in a discrete quantum system, the discrete Fourier transform can be used
to generate a second basis that is 
a mutually unbiased basis with respect to the first, and we use that basis to construct the momentum operator $\hat p$. 
In an orthonormal basis $\{ \ket{n}\}: n \in \mathbb{Z}_N$ for an $N$ dimensional discrete quantum space, 
the discrete Fourier transform  
\begin{align}
 \hat Q_{FT} &= \frac{1}{\sqrt{N}} \sum_{j,k=0}^{N-1} \omega^{jk}\ket{j}\bra{k},  
\end{align}
with 
\begin{align}
\omega &\equiv e^{2 \pi i/N}. 
\end{align}
Using the discrete Fourier transform, we construct an orthonormal basis consisting of states 
\begin{align}
\ket{m}_F  = \hat Q_{FT} \ket{m} 
= \frac{1}{\sqrt{N}} \sum_{j=0}^{N-1} \omega^{mj} \ket{j} 
\end{align}
with $m \in \mathbb{Z}_N$. 
Because
 \begin{align}
 \bra{j}\ket{m}_F = \frac{\omega^{jm}}{\sqrt{N}} .
 \end{align}
 has amplitude equal to $1/\sqrt{N}$,  independent of $j,m$, 
the basis  $\{ \ket{k}_F\}: k \in \mathbb{Z}_N$ is mutually unbiased
with respect to the basis $\{ \ket{n} \}: n \in \mathbb{Z}_N$. 

Because both $p, \phi \in [0, 2 \pi)$ and are both periodic in the classical Hamiltonian of equation 
\ref{eqn:Hclassical}, an operator for momentum, similar to that for the angle $\hat \phi$ (equation \ref{eqn:phihat}),
 can be defined in the Fourier basis 
\begin{align}
\hat p =\sum_{m=0}^{N-1} \frac{2 \pi m}{N} \ket{m}_F\bra{m}_F   . \label{eqn:phat}
\end{align}

With both dimensionless momentum and angle operators $\hat p, \hat \phi$ (equations \ref{eqn:phihat}, \ref{eqn:phat})
defined in a discrete quantum space, 
 the Hamiltonian of equation \ref{eqn:Hclassical} be written in terms of these operators
\begin{align}
\hat h(\tau)_{\rm PH} 
& =\hat h_0(\hat p,\hat \phi) + \hat h_1(\hat \phi, \tau)  \nonumber \\
\hat h_0(\hat \phi,\hat p) & =   a \left(1- \cos \hat p \right) - \epsilon \cos \hat \phi \nonumber \\
\hat h_1(\hat \phi, \tau)  &=  - \mu \cos(\hat \phi - \tau) - \mu' \cos(\hat \phi + \tau) . \label{eqn:Hquantum}
\end{align}

The operators $\hat p, \hat \phi$ don't commute, but  because our Hamiltonian
is separable, with kinetic energy term only a function 
of $\hat p$  and potential energy term only a function of $\hat \phi$, we don't need to consider the order of
these two operators during quantization.  
The eigenvalues of $\hat p$ and $\hat \phi$ are real but exhibit a jump of $2 \pi$ 
between $\ket{0}$ and $\ket{N-1}$ states and between $\ket{0}_F$ and $\ket{N-1}_F$ states, respectively.  
However, the jump in the eigenvalues of $\hat \phi$ and $\hat p$ does not affect the Hamiltonian operator 
of the Harper model because it is periodic in both operators.

To summarize, we quantize a periodic classical system that is described by 
a separable Hamiltonian in the following way that is similar to that used to study quantum systems on 
the torus (e.g., \citet{Saraceno_1990}). 
We choose a classical system with a Hamiltonian function that is a sum 
of a potential energy term that is a function of the canonical angle coordinate and a kinetic energy term
that is a function of the momentum. 
We restrict the phase space of the classical system so that it is doubly periodic.   
We select a discrete quantum space and create an operator for angle based on a basis labelled with integers.
We use the discrete Fourier transform to construct a Fourier basis and momentum operator that is
diagonal in the Fourier basis.  The Hamiltonian is converted to an operator using the position and momentum 
operators in the discrete quantum space.   This procedure gives us a map $\phi_N$, dependent upon 
positive integer $N$,  
from a classical doubly periodic and separable Hermitian operator (our classical Hamiltonian). 
In appendix \ref{sec:WigWeyl} we show how our quantization map is often equivalent  to that generated via a Wigner-Weyl transformation.

For the perturbed Harper model, the map between classical Hamiltonian and quantum  
operator is 
\begin{align}
H(\phi,p,t)_{\rm PH} \xrightarrow[\phi_N]{}  \hat h(\hat\phi,\hat p,t)_{\rm PH} 
\end{align}
with dimensionless classical Hamiltonian of equation \ref{eqn:Hclassical} on the left and 
the dimensionless Hermitian operator of equation \ref{eqn:Hquantum} on the right.

Because the discrete Fourier transform gives the basis of our momentum operator,
phase space contains $N$ states in both momentum and coordinate angle bases.
The discrete quantum space facilitates computation of quasi-energy levels and eigenstates as we
 can compute them using matrix operations, aiding in potential applications on systems that are 
 finite dimensional, such as quantum computers. 
 
 In odd dimensions, 
 a discrete version of Wigner-Weyl quantization that leverages discrete coherent state analogs generates 
 the same quantum system as discussed here, 
 and this is described in more detail in appendix \ref{sec:WigWeyl}. 

\subsection{Domain size and Planck's constant \label{sec:planck}}

While classical equations of motion use Hamilton's equations to generate the dynamics, 
in a quantum system with Hamiltonian $\hat H$, the time dependence 
 $\partial_t = - \frac{i}{\hbar} \hat H$  and is inversely proportional to the reduced Planck's
constant $\hbar$. From the dimensionless operator in Equation \ref{eqn:Hquantum}, we can 
construct a Hamiltonian operator with 
units of energy (and matching the classical Hamiltonian $K_{\rm PH}$ of equation \ref{eqn:Kclassical}) is  
\begin{align}
 \hat k_{\rm PH} = E_0 \hat h_{\rm PH} , 
\end{align}
 with $E_0 = L_0 \nu$. 

We apply the Bohr-Sommerfeld quantization condition $\int p dq = 2 \pi N \hbar$ 
(e.g., following \citet{Wei_1991} for the kicked quantum Harper model) 
where $N$ is a positive integer which we set 
to the dimension of our discrete quantum vector space. 
For our classical system, the momentum $p$ has units of angular momentum so the Bohr-Sommerfeld
quantization condition is $\int L d\phi  = 2 \pi n \hbar$.   The left hand size 
is the domain size in the phase space of our system which is equal to $(2 \pi)^2 L_0$. 
The Bohr-Sommerfeld  quantization condition gives 
\begin{align}
\frac{L_0}{\hbar} = \frac{N}{2 \pi} . \label{eqn:Lh}
\end{align}
With this prescription, 
the dimension of the discrete quantum space $N$ is 
related to the domain momentum size in units of $\hbar$. 
Equivalently if we choose $L_0$ and $N$, then $\hbar$ assumes a particular value. 

Equation \ref{eqn:Lh} implies that in the large $N$ limit, at a fixed value for $\hbar$, the size of the domain $ L_0 \to \infty$ 
and we would recover the perturbed pendulum model of equation \ref{eqn:PertPend}. 
We can also take the large $N$ limit at a fixed domain size $L_0$, which means that 
 $\hbar \to 0$, and is equivalent to the semi-classical limit. 
 Note $E_0 = \frac{N}{2 \pi} \hbar \nu$.   Again taking $E_0$ fixed, $N \to \infty$ is equivalent to 
 the classical limit $\hbar \to 0$.
Alternatively, $N \to \infty$ with $\hbar$ fixed gives $E_0 \to \infty$. 

The sensitivity of possible parameter values to $\hbar$ (and vice versa) when approximated
on a discrete quantum space also occurs 
with quantizations of the kicked rotor (e.g., \citet{Chirikov_1988}) and the kicked Harper model
(e.g., \citet{Wei_1991}).  

With the domain size in momentum spanning $[0, 2 \pi L_0]$, the operator 
$\hat p  = \frac{\hat L}{L_0}$.   Using equation \ref{eqn:Lh} 
the operator
\begin{align} 
\hat L  = \sum_m  \frac{2 \pi m}{N} L_0 \ket{m}\bra{m} = \sum_m \hbar m \ket{m}\bra{m},
\end{align} 
as expected.  Thus equation \ref{eqn:Hquantum} can also be written
\begin{align}
\hat h_{\rm PH} & =   a \left(1- \cos \frac{\hat L}{L_0} \right) - \epsilon \cos \hat \phi \nonumber \\
& \qquad - \mu \cos(\hat \phi - \tau) - \mu' \cos(\hat \phi + \tau) . 
\end{align}

For every classical system in the dimensionless form of equation 
\ref{eqn:Hclassical} we can associate a quantum system in the form of 
equation \ref{eqn:Hquantum}.  
To choose a discrete quantum system we must choose $N$, which via equation \ref{eqn:Lh}
sets  $L_0/\hbar  = N/(2\pi)$ and  $E_0/\hbar = L_0 \nu/\hbar = \nu N/(2 \pi)$. 

With $a, \epsilon >0$, 
the maximum energy $E_0(2a + \epsilon) $ and the minimum energy is $-E_0 \epsilon$.
With $N$ energy levels, the distance between the energy levels is about 
$\delta E \sim 2E_0 (\epsilon+a)/N$. 

Time evolution depends on the operator 
\begin{align}
 \frac{ \hat k_{\rm PH}(t)}{\hbar} = \frac{L_0 \nu \hat h_{\rm PH}(\tau)}{\hbar} = \frac{ N}{2 \pi} \nu \hat h_{\rm PH}(\tau)
 \label{eqn:ktoh}
 \end{align}
using equation \ref{eqn:Lh} and with $\tau = \nu t$. 
The above relation also implies 
\begin{align}
 \frac{ \hat k_{\rm PH} T }{\hbar}  = N \hat h_{\rm PH} 
\end{align}
which is relevant when computing the propagator for evolution over a duration equal to the perturbation period. 

The choice of a discrete quantum space facilitates numerical computation of the Floquet operator 
over the time $T = 2 \pi/\nu$, 
\begin{align}
\hat U_T & = {\mathcal{T}} e^{-\frac{i}{\hbar} \int_0^T \hat k( t)_{\rm PH} dt}   
\end{align}
where $\mathcal T$ indicates a product of time-ordered unitary operators computed in the limit
of small step size. 
With $n_t$ the number steps, and $dt = T/n_t$, 
\begin{align}
\hat U_T 
& =  \lim_{n_t \to \infty} {\mathcal T} \prod_{j=0}^{n_t-1} e^{- \frac{i}{\hbar} \hat k_{\rm PH}(j dt) \ dt}.
\end{align}
A statevector $\ket{\Psi(t)} $ that evolves via a periodic Hamiltonian operator 
such as  $\hat k(t)$ can be decomposed into a sum of Floquet states in the form 
$\ket{\psi_j(t)} = e^{i \alpha_j t} \ket{\phi_j(t)}$ where $e^{i \alpha_j T}$ are the eigenvalues of 
$\hat U_T$,  $\ket{\phi_j(t + T) }= \ket{\phi_j(t)}$ are periodic and $\ket{\phi_j(0)}$ are 
eigenstates of $\hat U_T$.  This decomposition is equivalent to the Floquet theorem. 

Using equation \ref{eqn:Lh} the Floquet propagator in terms of our rescaled 
Hamiltonian operator $\hat h_{\rm PH}$ (of equation \ref{eqn:Hquantum}) is 
\begin{align}
\hat U_T  & = {\mathcal T} e^{- \frac{i}{\tilde h} \int_0^{2 \pi} \hat h(\tau)_{\rm PH} d \tau } 
 \label{eqn:U_T} 
\end{align} 
with 
\begin{align}
\tilde h = \frac{2 \pi}{N} 
\end{align}
acting like $\hbar$ in our rescaled system.  
Planck's constant is similarly rescaled 
when quantizing the kicked rotor \citep{Chirikov_1988}. 
The unitary operator $\hat U_T$ describes the evolution of 
 our rescaled quantum system from dimensionless time $\tau=0$ to $2 \pi$. 

Parameters describing  the quantum system (equation \ref{eqn:Hquantum}) 
include those chosen for the associated classical system 
 (equation \ref{eqn:Hclassical}); 
 $a, \epsilon, \mu, \mu'$, also listed in equation \ref{eqn:dofs}). 
In addition to those, we also choose the  dimension $N$ of the associated discrete quantum space which 
 in turn sets the dimensionless ratio $L_0/\hbar$.  
 
\begin{figure*}[htbp]\centering
\includegraphics[width=2.1 truein, trim=15 15 10 10,clip]{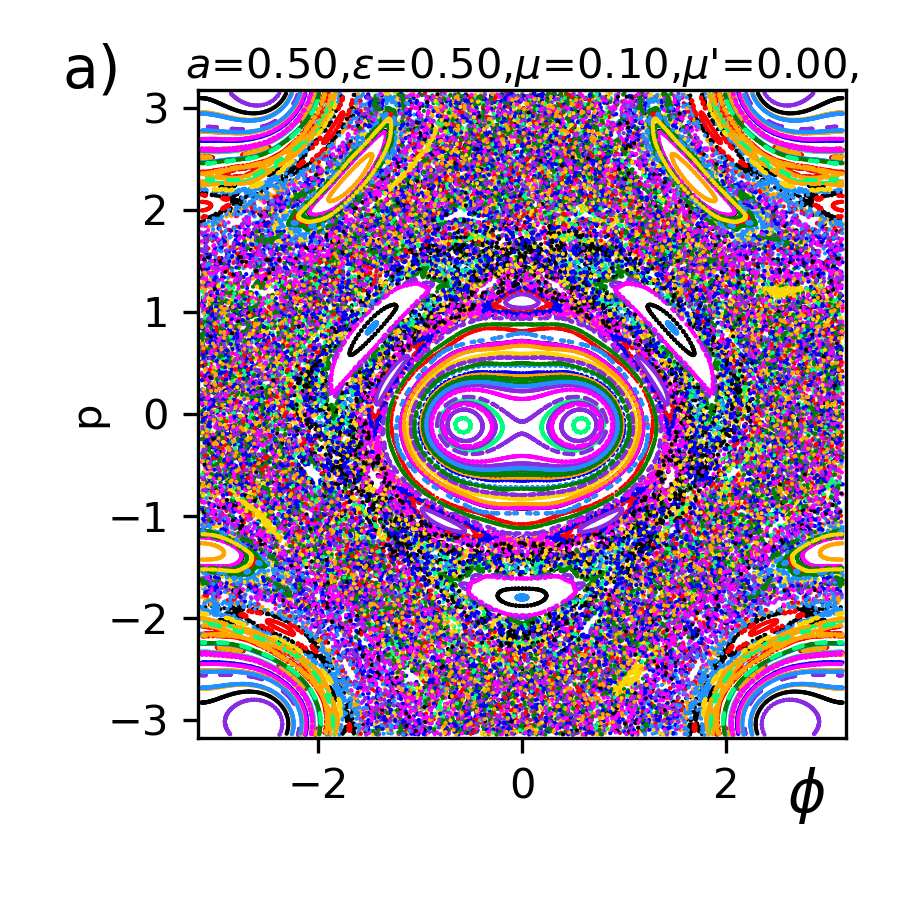}
\includegraphics[width=2.1 truein,trim = 0 -10 0 0,clip]{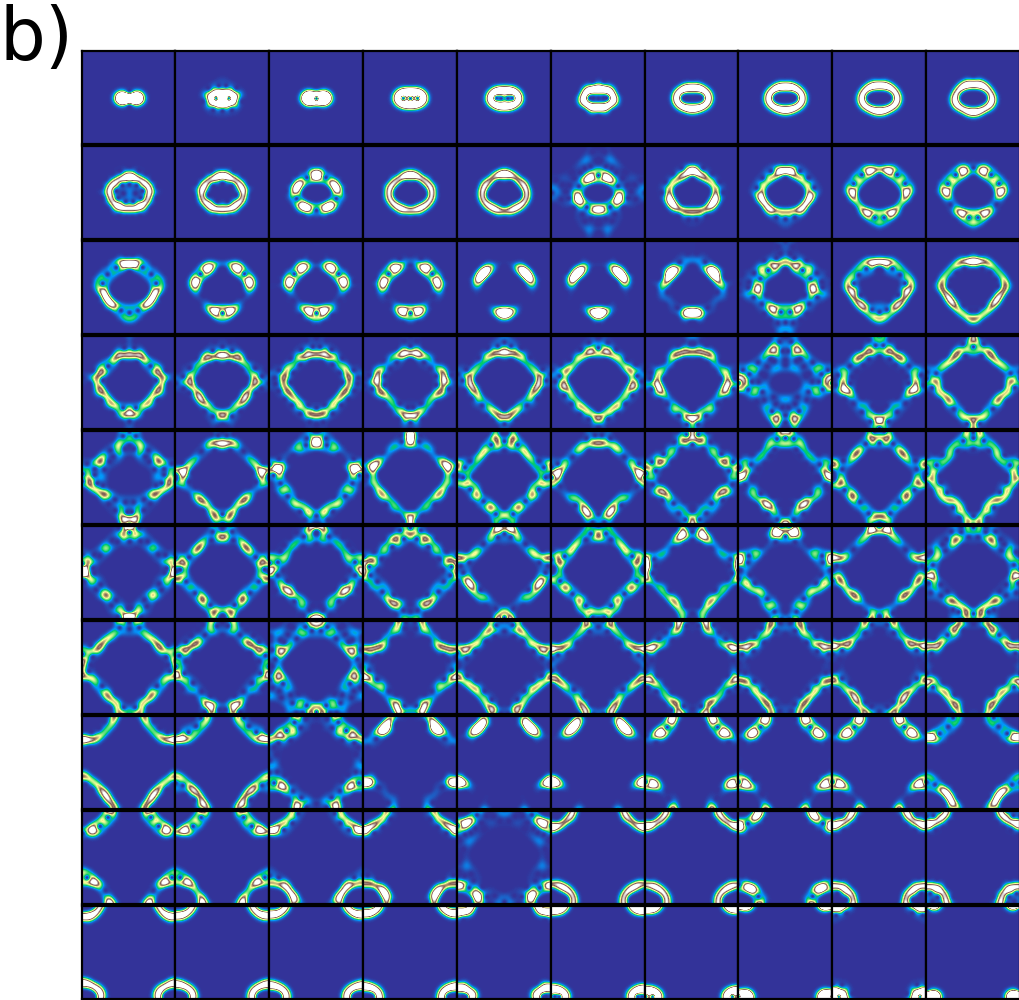}
\includegraphics[width=2.3 truein,trim = 0 0 0 0,clip]{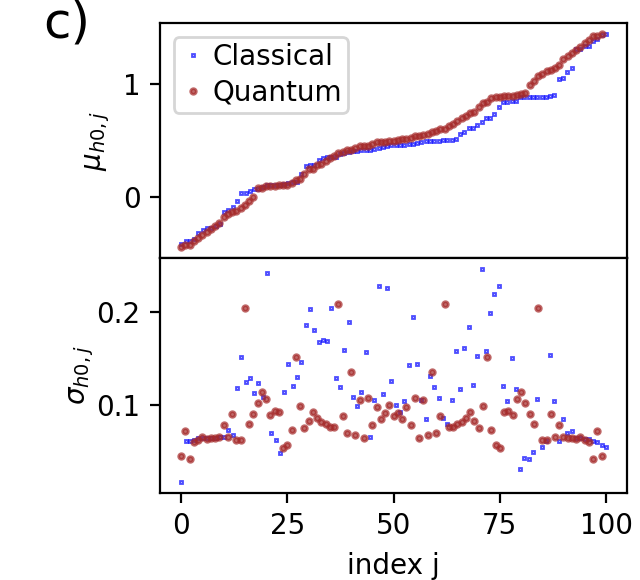}
\caption{Similar to Figure \ref{fig:Q9b} except for a classical system that has larger chaotic regions
and has an asymmetric perturbation; $\mu \ne \mu'$.  
This system also has $N=100$ states. 
\label{fig:Q9c}}
\end{figure*}

\begin{figure*}[ht]\centering
\includegraphics[width=5 truein,trim = 0 0 0 15,clip]{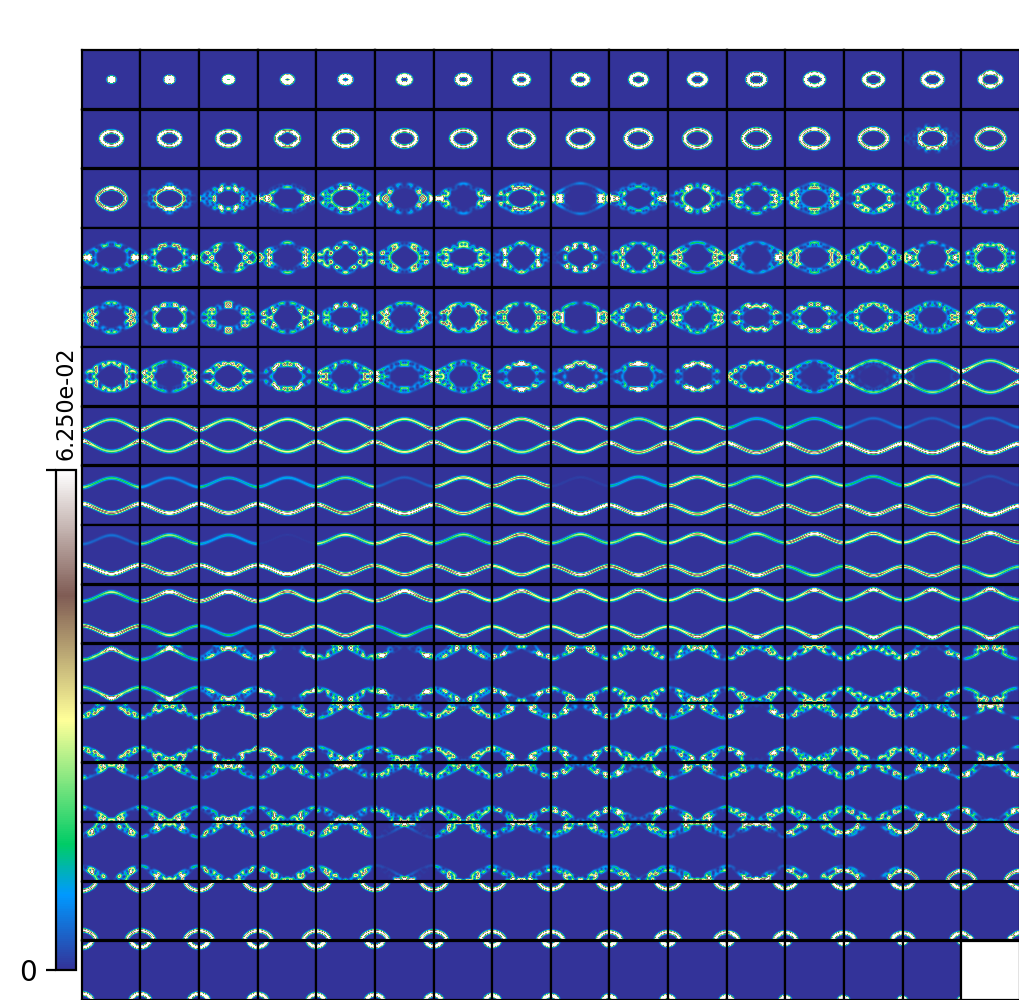}
\caption{Similar to Figure \ref{fig:Q9b}b except that the number of states $N=255$ instead of 100. 
The Hamiltonian has the same parameters $a,\epsilon, \mu, \mu'$ as 
 the system shown in  Figure \ref{fig:Q9b}. 
The Husimi distributions are similar with larger $N$, though there may be additional 
localized states located with the chaotic region.  
 \label{fig:Q9b_large}}
\end{figure*}

\section{Numerical computations \label{sec:num}}

Using the procedure for Suzuki-Trotter decomposition described in appendix \ref{sec:trotter}, we compute 
the propagator $\hat U_T$ for some quantized periodically perturbed Harper models 
(with Hamiltonian of equation \ref{eqn:Hquantum}) and compare them to their associated classical model
(with Hamiltonian of equation \ref{eqn:Hclassical}).    
Three different model systems are shown in Figures \ref{fig:vanil} -- \ref{fig:Q9c}.  
We show orbits of the associated classical model in the a) panels,  and in the  b) panels we show Husimi distributions, computed as described in appendix \ref{sec:Hus},
  for the eigenvectors of the propagator $\hat U_T$.    
The center of the a) subfigures is the origin $(\phi,p) = (0, 0)$.  Each panel in 
the b) subfigure shows an $N\times N$
array computed with a single eigenstate of the Floquet propagator $\hat U_T$. 
Each pixel of a panel has different indices $k,l$ setting the coherent 
state for computing equation \ref{eqn:Hus}, and with 
the central value in the image corresponding has indices $k,l=0$ so as to match the $(\phi,p) =(0,0)$  
central position of the classical phase space surfaces of section.   This makes it possible to 
visually compare the classical phase space surfaces of sections in the a) subfigure to the Husimi distributions 
in the b) subfigure. 
In Figures \ref{fig:vanil}, \ref{fig:Q9b} and \ref{fig:Q9c}, the dimension of the quantum system  $N=100$. 
The Husimi distributions shown in Figure \ref{fig:Q9b_large} have the same parameters 
$a,\epsilon, \mu, \mu'$ as  the system shown in Figure \ref{fig:Q9b} but the dimension is larger, $N=255$. 
 The Husimi distributions are displayed with a maximum level of $1/\sqrt{N}$ and a colorbar is included on 
the left sides of Figure \ref{fig:vanil}b and Figure \ref{fig:Q9b_large}.   
In Figures \ref{fig:Q9b}b and \ref{fig:Q9c}b, with dimension $N=100$,  
the colorbar is the same as shown in Figure \ref{fig:vanil}b.

A comparison between the Husimi distributions and the classical orbits seen in the surfaces of section
 illustrates that the eigenstates of 
the quantized models are closely related to the orbits of the classical system.  
For the systems shown in Figures \ref{fig:Q9b} and \ref{fig:Q9c} the phase space regions containing  
 area filling chaotic classical orbits, have Husimi distributions that are wider and cover the chaotic region. 
Surfaces of section and associated Husimi distributions generated starting at an intermediate time 
within the driving period are shown in appendix \ref{sec:shift}. 
 
\subsection{An order for the eigenstates of the Floquet propagator}

The Floquet propagator, $\hat U_T$ (equation \ref{eqn:U_T}), because it is unitary, has complex eigenvalues of magnitude 1. 
An eigenvalue of $\hat U_T$ in the form $e^{i \lambda}$  
can be described by its argument or phase, real $\lambda$, which can be called a  quasi-energy.
We find that the Husimi distributions of the eigenvectors of $\hat U_T$ appear to 
be in a random order if we sort them in order of increasing quasi-energy.  
Instead, we sort the eigenstates of $\hat U_T$ in order of their expectation value of the unperturbed 
and static portion of the Hamiltonian,  $\hat h_0$ (specified in equation \ref{eqn:Hquantum}). 
For each eigenstate $\ket{w_j}$ (indexed by $j$) of $\hat U_T$, we compute the expectation value
\begin{align}
\mu_{h0,j} \equiv \bra{w_j} \hat h_0 \ket{w_j}. \label{eqn:muh0}
\end{align}
Our approach to ordering eigenstates of the propagator 
is similar to \citet{Le_2020} but we use the static and unperturbed energy instead 
of the time average of the energy. 
 
Figures \ref{fig:vanil}b, \ref{fig:Q9b}b, and \ref{fig:Q9c}b show the Husimi 
distributions for the eigenstates of $\hat U_T$ in consecutive order of expectation value $\mu_{h0,j}$. 
We find that the Husimi distribution of an eigenstate $\ket{w_j}$ 
of the Floquet propagator $\hat U_T$ resembles classical orbits of the associated classical system with energy near $\bra{w_j} \hat h_0 \ket{w_j}$. 

\subsection{Energy dispersions and ergodicity \label{sec:disp}}

In the top panels of Figures \ref{fig:vanil}c, \ref{fig:Q9b}c, and \ref{fig:Q9c}c we plot with brown dots 
the expectation $\mu_{h0,j}$ of the unperturbed Hamiltonian $\hat h_0$ for each eigenstate $\ket{w_j}$ of the Floquet propagator $\hat U_T$.  
In the bottom panels we plot the standard deviation $\sigma_{h0,j}$ for each eigenstate, where 
\begin{align}
\sigma_{h0,j} = \sqrt{\bra{w_j} \hat h_0^2 \ket{w_j} - (\bra{w_j} \hat h_0\ket{w_j})^2 }.
\label{eqn:sigh0}
\end{align}  
The means and standard deviations are plotted
as a function of index $j$ after putting the states in order of increasing $\mu_{h0,j}$. 
Figures \ref{fig:Q9b}c, and \ref{fig:Q9c}c show that where classical orbits are chaotic, area filling,  and 
exhibit variations in the energy of the unperturbed Hamiltonian function, the quantum eigenstates 
have a higher standard deviation $\sigma_{h0,j}$. 
For our perturbed Harper system, the standard deviation $\sigma_{h0,j}$, 
computed from the eigenstates of the Floquet propagator, measures of the width (in energy) of
the chaotic region. 

In Figures \ref{fig:vanil}c, \ref{fig:Q9b}c, and \ref{fig:Q9c}c and 
for their associated classical Hamiltonian, 
we integrate for 400 periods,  giving a series of 400 points (each separated in time by a 
period) in phase space, for each of 500 orbits. 
The initial conditions for these 500 orbits are randomly chosen from 
a uniform distribution that covers phase space.  
The 400 phase space positions from each orbit give a series of values for the energy $H_0(\phi,p)$, the value of the classical Hamiltonian.  
The mean $\mu_{H0}$ and standard deviation $\sigma_{H0}$ for the unperturbed Hamiltonian function $H_0(p,\phi)$ are then computed from the points in each orbit.
The orbits are sorted in order of their mean energy, $\mu_{H0}$,  
and these mean values shown in Figures \ref{fig:vanil}c -- \ref{fig:Q9c}c top panel with blue squares. 
The bottom panels in Figure \ref{fig:vanil}c -- \ref{fig:Q9c}c  show the standard deviations $\sigma_{H0}$ computed from the same orbits. 
While the horizontal axis on this plot is given with respect to the index of the sorted eigenstates of 
the quantum system, we have displayed the classically computed energy means and standard deviations 
so as to cover the same horizontal range.  We don't expect the quantum means and standard deviations to exactly match the classical ones
because the classical orbit initial conditions were randomly chosen. 
Figures \ref{fig:Q9b}c, and \ref{fig:Q9c}c illustrate that the means and standard deviations  
of the unperturbed Hamiltonian classically computed from the orbits 
are similar to the means and standard deviations computed from the from eigenstates of the associated quantum model.  The similarity between quantum and classical energy dispersions suggests
that the dispersion of the unperturbed energy gives an quantitative estimate for 
ergodicity in both settings.  

Figure \ref{fig:Q9b_large} shows a quantum system identical to that shown in Figure \ref{fig:Q9b}b 
except that the number of states $N=255$ instead of 100.    
The similarity of the Husimi function at $N=255$ and $N=100$ illustrates that the 
ergodicity is not reduced at larger $N$. 
In the $N=255$ system, there is a localized eigenstate
embedded within the chaotic zone (in about the middle of the third row from the top).   
This behavior is perhaps similar to the phenomena of quantum scars seen in billiard systems and 
near periodic classical orbits \citep{Heller_1984,Berry_1989}.   
We chose an odd dimension for this system and checked that the properties of 
 the Floquet propagator eigenstates (such as the appearance of the Husimi distributions) are not strongly dependent upon whether the dimension $N$ is even or odd. 
 
In Figures \ref{fig:Q9b}c and \ref{fig:Q9c}c, the classical orbits within the chaotic region have similar average
energy, (and had we integrated the orbits longer, the average could have approached a single value). 
The differences between energies of the  eigenstates (computed via an expectation value of $\hat h_0$; equation 
\ref{eqn:muh0}) in this chaotic region are reduced. 
In the high $N$ limit, corresponding to the classical limit, we expect that the differences
between the energies of the eigenstates would be even smaller. 
 
In a classical system, one notion of 
ergodicity is that the time average of
an observable is equal to its spatial average. 
This notion was extended to
 quantum systems by \citet{Shnirelman_1974}.   When the classical dynamics is ergodic, for almost 
all eigenstates, the expectation values of observables converges to the phase-space average
 in the limit the dimension of the quantum space goes to infinity,  $N \to \infty$ 
 (also see \citet{Kurlberg_2001}).   

The close correspondence between the energy dispersions computed from the classical 
orbits and the quantum equivalent computed from the eigenstates of the Floquet propagator
(shown in in Figures  \ref{fig:vanil}c, \ref{fig:Q9b}c, and \ref{fig:Q9c}c) 
suggests that the dispersion of the 
unperturbed Hamiltonian (equation \ref{eqn:sigh0}) gives a quantitative estimate for 
quantum ergodicity.   For two systems with the same values of parameters $a, \epsilon,\mu, \mu'$ but with two different dimensions $N$  (Figures \ref{fig:Q9b}b and \ref{fig:Q9b_large}b)
 we find that the eigenstates of the Floquet propagator that are 
within the classical chaotic region in phase space have similar high standard deviations 
$\sigma_{h0,j}$ in both cases. 
This suggests that these quantum systems would be ergodic in the sense of a high $N$ limit, 
(following \citep{Shnirelman_1974,Kurlberg_2001}.  
At larger $N$ there could still be a small subset of 
eigenstates of the Floquet propagator within the chaotic regions, that are strongly localized  
or associated with periodic orbits,  akin to quantum scars (e.g., \citep{Heller_1984,Berry_1989}).

For the mixed systems shown in Figures \ref{fig:Q9b} and \ref{fig:Q9c}, the Husimi functions 
exhibit a remarkable morphological change at the chaotic region boundary, with  
ergodic eigenstates associated with ergodic orbits filling a larger region in phase space than 
neighboring eigenstates which resemble integrable classical orbits.  
Similar behavior was noted by \citet{Wang_2023} near the chaotic boundary for the quantized kicked rotor. 

\begin{figure*}[htbp]\centering
\includegraphics[width=5.5 truein, trim=0 0 0 0,clip]{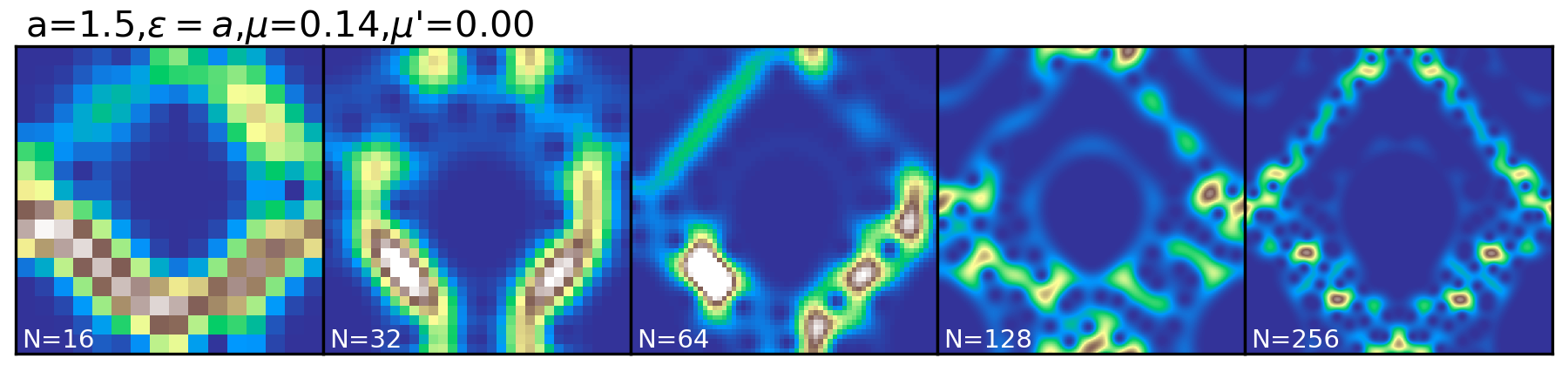}
\caption{A sequence of Husimi distributions as a function of increasing dimension $N$.
We show Husimi functions constructed from eigenstates of the Floquet propagator $\hat U_T$ that have energy near the separatrix for a Hamiltonian model with parameters 
 printed on the top left side of the plot.  
Each panel shows a model  that has the same parameters but has a different dimension $N$. 
The dimensions are printed on the bottom left of each panel and  are powers of 2. 
This figure illustrates that in a region that is chaotic in the associated classical model, 
Husimi functions of the eigenstates appear increasingly 
diffuse and evenly distributed at larger $N$.  This suggests that they belong to a subspace 
that is ergodic in the sense described by \citet{Shnirelman_1974} (also see \citep{Kurlberg_2001}). 
\label{fig:ergo}
}
\end{figure*}

To illustrate the notion of ergodicity in the quantum system,  
in Figure \ref{fig:ergo} we show a series of Husimi distributions for Hamiltonian 
models that have the same parameters except each one has a different value of $N$.
From left to right,  the dimension increases by a factor of 2 from panel to panel. 
Each Husimi distribution is constructed from an eigenstate of the Floquet propagator that has
 energy near the separatrix.  We choose $a=\epsilon$ so that there is only a single
 separatrix but set $\mu' \ne \mu$ so the model is lopsided like that shown in Figure \ref{fig:Q9c}.
The separatrix region is chaotic and at larger $N$ the eigenstates appear increasingly diffuse 
and evenly distributed.  This suggests that in the high $N$ limit  regions that 
are chaotic in the classical system could correspond to an ergodic subspace that is ergodic in the
sense defined by \citet{Shnirelman_1974,Kurlberg_2001}.

To provide context with prior studies (e.g., \citep{Haake_2010,Cohen_2023}) of quasi-energy spacing distributions for other quantized Hamiltonian chaotic systems,  we show in appendix \ref{sec:quasi} 
 quasi-energy spacing distributions for non-chaotic and ergodic systems and 
for non-chaotic and ergodic subspaces of a mixed system. 

\section{Estimates for the width of the chaotic region} 

\subsection{The width of the classical chaotic region at the separatrix \label{sec:MA}}

In a classical Hamiltonian system, a method akin to 
Melnikov's method for proving existence of chaotic behavior \citep{Melnikov_1963} 
can be used to estimate the size of the energy change caused
by a perturbation near the separatrix orbit of the unperturbed system 
\citep{Zaslavsky_1968,Chirikov_1979,Shevchenko_2000,Treschev_2010}. 
The width of the chaotic region formed at the separatrix connecting two
hyperbolic fixed points is estimated by integrating the time dependent perturbation along the separatrix 
orbit that is present in the unperturbed system.  The change in energy caused by the perturbation on the 
the separatrix orbit with momentum and angle $p_s(t), \phi_s(t)$ is  
\begin{align}
\Delta H = \int \frac{\partial H_1(\phi_s,p_s, t)}{\partial t } dt.   \label{eqn:DeltaH_C}
\end{align}

When applied to a perturbed pendulum, (equation \ref{eqn:PertPend}) the energy change 
$\Delta H$ depends upon the Melnikov-Arnold integral \citep{Chirikov_1979,Lichtenberg_1992,Shevchenko_2000}
 \begin{align}
A_m(\lambda) &\equiv \int_{-\infty}^\infty \cos( \frac{1}{2} m \phi_s(t) -\lambda t )\ dt  \label{eqn:MA}
\end{align}
on the separatrix trajectory 
\begin{align}
\phi_s(t) &= 4\ {\rm arctan} (e^t) - \pi  . \label{eqn:traj}
\end{align}
For $\lambda<0$, the result is typically small.  For $\lambda >0$  (and following 
\citep{Chirikov_1979,Lichtenberg_1992})
\begin{align}
A_1(\lambda) & = 2 \pi \frac{e^\frac{\pi \lambda}{2 }}{\rm {sinh} (\pi \lambda)} \label{eqn:A_1} \\ 
A_2(\lambda) & = 2 \lambda A_1(\lambda) . \label{eqn:A_2}
\end{align}
Following \citet{Shevchenko_2000}, 
for the particular  pendulum system $H_0(p,\phi) = \frac{p^2}{2} - \omega_0^2 \cos \phi$ 
with perturbation $\mu \cos \phi\cos \tau$, 
the separatrix trajectory is that of equation \ref{eqn:traj} but with time rescaled by $\omega_0$.  
The half width of the chaotic region near the separatrix is estimated as 
\begin{align}
\Delta H & \sim \frac{\mu}{\omega_0} (A_2(\omega_0^{-1}) + A_2(-\omega_0^{-1})) \nonumber \\
&\sim  \frac{4 \pi \mu}{\omega_0^2} \frac{ e^{\frac{\pi}{2 \omega_0}} } { {\rm sinh} ( \pi/\omega_0)} .
 \label{eqn:DeltaH0}
\end{align}
The perturbation frequency is absent because we work with time in units of the perturbation frequency. 
Our factor 
$1/\omega_0$ represents the ratio of perturbation to pendulum libration frequency. 
Here we have neglected a relative phase shift for the perturbation, but it is taken into account 
when generating separatrix maps \citep{Zaslavsky_1968,Chirikov_1979,Lichtenberg_1992,Soskin_2009}.   
Better approximations to the separatrix width 
take into account a series of strong  resonances within the separatrix chaotic layer 
and are usually calculated in the weak perturbation limit of $\mu/\epsilon \ll 1$ \citep{Shevchenko_2008,Soskin_2009}. 

Using an integral along the separatrix orbit, 
we similarly estimate the width of the chaotic region 
for special cases of the periodically perturbed Harper model of equation \ref{eqn:Hclassical}.

We estimate the width of the primary resonance using the energy of the separatrix orbit of the 
unperturbed system with energy $\epsilon$, assuming  $\epsilon, a >0$. 
The maximum momentum of this orbit satisfies 
$p_{\rm max} = 2 {\rm arcsin} \sqrt{\epsilon/a}$ which is narrow if $\sqrt{\epsilon/a}$ is small. 
In the limit of small $\sqrt{\epsilon/a}$, 
the primary resonance resembles a pendulum with $H_0(p,\phi) \sim a p^2/2 - \epsilon \cos \phi$. 
With perturbation $-\mu( \cos (\phi-\tau) + \cos (\phi + \tau) )$, 
the energy change at the separatrix is that given by equation  \ref{eqn:DeltaH0} 
but with $\omega_0 = \sqrt{\epsilon/a}$. 

In the case of $a = \epsilon$, the energy of the two hyperbolic fixed points are the same and there is a single separatrix, otherwise there are two separatrices that 
 separate phase space into 3 distinct regions,  as seen in Figure \ref{fig:vanil}a. 
The separatrix satisfies
$\cos p = - \cos \phi$ or $p = \pm \phi + \pi$.  
Hamilton's equations give 
$\dot \phi  = \partial_p H_0 = a \sin p$.  On the separatrix 
orbit  $\dot \phi = \pm a \sin \phi $ and  
 $d\tau =\pm \frac{d\phi}{a\sin \phi}$.  
We integrate to find a separatrix orbit 
\begin{align}
\phi_s(\tau) &= 2\ {\rm arctan}( e^{\pm a \tau} )
 . \label{eqn:sep_orbit}
\end{align}  
The sign specifies the particular segment of the separatrix. 
The trajectory resembles that of equation \ref{eqn:traj} for the pendulum's separatrix except for a factor of 2.
The energy change for $\mu = \mu'$ 
depends upon the Melnikov-Arnold integral $A_1()$ instead of $A_2()$ giving half width energy estimate 
\begin{align}
\Delta H \sim  \frac{2 \pi \mu}{\omega_0} \frac{e^{\frac{\pi}{2 \omega_0}} }{{\rm sinh}  {\pi /\omega_0}} , 
 \label{eqn:DeltaH0s}
\end{align}
with $\omega_0 = a$ because $a = \epsilon$ for this  case. 
A comparison between equation \ref{eqn:DeltaH0s} and equation \ref{eqn:DeltaH0}
suggests that a general expression could be developed that interpolates between these cases.
Though we found simple analytical forms for 
the separatrix orbits with $a \ne \epsilon$  (equations \ref{eqn:sep_orbits} can be 
integrated to give expressions for $\phi_s(\tau)$),  we have not 
found a simple analytical form for the integral along the separatrix in the more general case of the Harper model. 

We computed the width of the chaotic region numerically from the energy dispersion 
for models with  $a = \epsilon$, $\omega_0 \sim 1$,  and $\mu $ ranging from 0.01 to 0.1 
and find that the chaotic region width 
estimated by equation \ref{eqn:DeltaH0s} is a factor of a few lower than we measured.  
This is not inconsistent with more detailed and improved estimates for the chaotic zone width 
that are predominantly tested in the 
weak perturbation limit of $\mu/\epsilon \ll 1$ \citep{Shevchenko_2008,Soskin_2009}. 

\subsection{Averaging the Floquet propagator in the interaction representation \label{sec:ave}}

We derive an expression for $\Delta H$, the width of the chaotic region and analogous to 
equation \ref{eqn:DeltaH_C}, but in the quantum regime. 
For some time dependent Hamiltonian systems, it can be 
efficient to compute a propagator using the interaction representation (or picture) as 
the Magnus expansion
in this representation is more likely to converge 
(e.g., \citep{Brinkmann_2016,Low_2018,Chen_2021}). 
In the interaction representation, 
with Hamiltonian operator $\hat h = \hat h_0 + \hat h_1(t)$,  
a sum of time independent and time dependent operators, state vectors 
are defined with a time-dependent unitary transformation 
\begin{align}
\ket{\psi_I(t)} = e^{i \hat h_0 t/\hbar} \ket{\psi(0)} .
\end{align}  
The operator associated with perturbation operator $\hat h_1(t)$  is 
\begin{align}
\hat h_{1,I}(t) = e^{\frac{i}{\hbar} \hat h_0 t} \hat h_1(t) e^{-\frac{i}{\hbar}\hat h_0 t} . \label{eqn:h_1I}
\end{align}
For our quantum system with Hamiltonian in equation \ref{eqn:Hquantum},
 with $\hbar = 2\pi/N$ (as discussed in 
section \ref{sec:planck} for our dimensionless Hamiltonian operator, see equation \ref{eqn:U_T}), 
\begin{align}
\hat h_{1,I}(\tau) &=  e^{\frac{iN}{2\pi} \hat h_0 \tau} \hat h_1(\tau) e^{-\frac{iN}{2 \pi}\hat h_0 \tau} 
\label{eqn:h1_first} \\
&= \sum_{j=0}^\infty \left( \frac{iN\tau}{2\pi} \right)^j\frac{1}{j!}  [(\hat h_0)^j, \hat h_1 (\tau)] \label{eqn:h1_I}
\end{align}
where in the second line we have 
used the Campbell identity \citep{Campbell_1897} and short hand notation 
for repeated commutators (e.g., $[(\hat h_0)^3, \hat h_1] = [\hat h_0, [\hat h_0, [\hat h_0, \hat h_1]]]$
and $[(\hat h_0)^0, \hat h_1]   =\hat h_1$).

In the interaction representation,  the Floquet propagator of equation \ref{eqn:U_T} becomes
\begin{align}
\hat U_{T,I} = {\mathcal T} e^{\frac{Ni}{2\pi} \int_0^{2 \pi}  \hat h_{1,I}(\tau) d\tau  }. \label{eqn:UTI}
\end{align}

The Magnus expansion gives an 
 time-independent effective or average Hamiltonian $h_{F,I}$, for a Floquet propagator such as in equation 
 \ref{eqn:UTI}, \citep{Magnus_1954,Chu_1985,Blanes_2009,Brinkmann_2016}
 \begin{align}
 e^{-iN \hat h_{F,I}} = U_{T,I}. \label{eqn:h_F}
 \end{align}
The expansion is the sum of a series of operators $\hat \Omega_n$ 
\begin{align}
N\hat h_{F,I} &= \sum_{n=1}^\infty \hat \Omega_n ,  \label{eqn:hFI_def}
\end{align}
and the first two terms in the expansion are 
\begin{align}
\hat \Omega_1 &=\frac{N}{2\pi} \int_0^{2 \pi} \hat h_{1,I}(\tau) d\tau \label{eqn:Om1_a} \\
\hat \Omega_2 &= \left( \frac{N}{2\pi}\right)^2 \frac{1}{2i} \int_0^{2 \pi}  d\tau_1 \int_0^{\tau_1} d\tau_2
 [\hat h_{1,I}(\tau_1) ,\hat h_{1,I}(\tau_2) ].  \label{eqn:Om2}
\end{align}
The expansion must converge if
\begin{align}
\frac{N}{2\pi} \int_0^{2 \pi} \lVert  \hat h_{1,I}(\tau) \rVert_2 d\tau  < \pi  \label{eqn:convergence}
\end{align}
where $\lVert. \rVert_2$ denotes the matrix norm. 
In this section 
we adopt a perturbation operator (from equation \ref{eqn:Hquantum}) with $\mu = \mu'$ 
\begin{align}
\hat h_1 (\tau) = - \mu \cos \hat \phi \cos \tau  \label{eqn:ourpert}
\end{align}
so that we can describe the strength of the time dependent perturbation with a single parameter, $\mu$. 
With this perturbation 
\begin{align}
\frac{N}{2\pi} \int_0^{2\pi} \lVert \hat h_{1,I} (\tau) \rVert_2 d\tau \le N |\mu|.
\end{align}
If the perturbation parameter $\mu$ is small enough, the Magnus expansion of the
interaction propagator (equation \ref{eqn:UTI}) would definitively converge. 
We note that the convergence condition is not fulfilled in many practical applications of
the Magnus expansion (e.g., \cite{Blanes_2009,Kuwahara_2016}). 

The first operator, $\hat \Omega_1$ in the Magnus expansion propagator in the 
interaction representation, is first order in the perturbation parameter $\mu$.  
Inspection of equation \ref{eqn:Om2} and the interaction Hamiltonian of 
equation \ref{eqn:h1_I} implies the order of $\mu$ is equal to the index of the 
operator in the expansion. 
The averaged Hamiltonian 
\begin{align}
\hat h_{F,I} &\approx \frac{1}{2\pi} \int_0^{2 \pi} d \tau \ \hat h_{1,I} (\tau) + { \mathcal O} (\mu^2) .
\label{eqn:FF}
\end{align} 
Equation \ref{eqn:FF} shows that $\hat h_{F,I}$ is the average of the perturbation 
in the interaction representation and explains why we included a factor of $N$ in the exponent on 
the lefthand side of equation \ref{eqn:hFI_def} in our definition for $\hat h_{F,I}$. 
This equation suggests that the averaged Hamiltonian in the interaction representation could give a 
description for the separatrix width, similar to the integral of the perturbation along the separatrix 
in the classical setting which is also first order in perturbation parameter $\mu$.  

We compute $\hat h_{F,I}$ to first order in $\mu$ (using equation \ref{eqn:h1_I})
\begin{align}
\hat h_{F,I}
& =\frac{1}{2\pi} \int_0^{2 \pi} d\tau \sum_{j=0}^\infty \left( \frac{iN\tau}{2\pi} \right)^j\frac{1}{j!}  [(\hat h_0)^j, \hat h_{1} (\tau)].
\end{align}
The $j=0$ term vanishes as 
as long as $\int_0^{2\pi} \hat h_1(\tau) = 0$, and this is obeyed by the perturbation in our model of equation \ref{eqn:Hquantum}. 
The $j=1$ term vanishes because $\int_0^{2\pi} \tau \cos \tau d \tau = 0$. 
The first non-vanishing term is that with $j=2$. 
For $j \ge 2$ we can integrate the integral $\int_0^{2 \pi} \tau^j \cos \tau d\tau $ to find a general expression 
for the operator 
\begin{align}
\hat h_{F,I} &=- \frac{\mu }{2\pi} \sum_{j=2}^\infty [(\hat h_0)^j, \cos \hat \phi] \left( \frac{iN}{2\pi} \right)^j \nonumber \\
& \times
 \Bigg(\sum_{k=0}^j  
 \frac{ (2 \pi)^{j-k}  }{(j-k)!}  \sin \frac{k\pi}{2}   -  \sin \frac{j\pi}{2} \Bigg). \label{eqn:Om1}
\end{align}
In appendix \ref{sec:clock_shift} we computed the commutators 
$[\cos \hat p, \cos \hat \phi]$ and $[\cos \hat p, \sin \hat \phi]$
(equations \ref{eqn:coms}) and found that they were approximately proportional to $2 \pi/N$. 
The factors of $N$ would tend to cancel for each term indexed by $j$ in equation \ref{eqn:Om1}.    
Probably the terms decrease in size, as the interaction  
Hamiltonian is bounded, but they might not decrease in size very quickly. 
We found that the eigenstates of $\hat h_0$ that are associated with stable fixed points in the 
associated classical model  
are approximately eigenstates of $\cos \hat \phi$ and so do not contribute to $\hat h_{F,I}$. 
The commutators in equation \ref{eqn:Om1} determine which eigenstates (of $\hat h_0$) are most 
affected by the time dependent perturbation. 
 
We denote an eigenstate of the unperturbed Hamiltonian $\hat h_0$ with energy $E_j$ as  $\ket{v_j}$ 
and use shorthand 
\begin{align} 
E_{jk} = E_j - E_k. 
\end{align}
We compute the matrix values for $\hat h_{F,I}$ using the eigenvectors of $\hat h_0$ in 
the interaction representation. 
The eigenstates of $\hat h_0$ in the interaction representation 
\begin{align}
\ket{v_{j,I}(\tau)} = e^{ \frac{iN}{2\pi} E_j \tau}\ket{v_j} .  
\end{align}
Using the states $\ket{v_{j,I}(\tau)}$, 
we take matrix values for $\hat h_{F,I}(t)$ from equation \ref{eqn:h1_first}, 
insert the perturbation from equation \ref{eqn:ourpert}, and  then integrate over the perturbation period as shown in equation \ref{eqn:FF}.   This gives 
matrix values for $\hat h_{F,I}$ (in the interaction picture) 
\begin{align}
\bra{v_{j,I}(\tau)} \hat h_{F,I} \ket{v_{k,I} (\tau) }
& \approx \frac{\mu i}{2\pi} (e^{ i N E_{jk} } -1)  \frac{NE_{jk}}{2\pi} 
\frac{\bra{v_j} \cos \hat \phi \ket{v_k}  } { \left( \frac{NE_{jk}}{2\pi} \right)^2 - 1}\nonumber \\
& \qquad    \times e^{-\frac{i}{2\pi} NE_{jk} \tau}
. 
\label{eqn:mat_I}
\end{align}

Because we are working in the interaction picture, 
equation \ref{eqn:h_F} implies that $\hat h_{F, I}$ is  a perturbation 
to the Hamiltonian $\hat h_0$. 
In the Schr\"odinger picture, the Floquet propagator  
can be written in terms of a constant averaged Hamiltonian plus a perturbation  $\mu \hat V $
evolved across a period of $2 \pi$, 
\begin{align}
\hat U_T = e^{- iN( \hat h_0 + \mu \hat V)}. \label{eqn:hV}
\end{align} 
Consequently, the matrix elements of the perturbation $\mu \hat V$ are equivalent to those computed 
in equation \ref{eqn:mat_I} via $\hat h_{F,I}$ at $\tau =2 \pi$, 
\begin{align}
V_{jk} &= \bra{v_j} \hat V \ket{v_k}  = \frac{1}{\mu} \bra{v_{j,I}(2\pi)} \hat h_{F,I} \ket{v_{k,I}(2\pi)} \\
& = \frac{ i}{2\pi} (e^{ i N E_{jk} } -1)  \frac{NE_{jk}}{2\pi} 
\frac{\bra{v_j} \cos \hat \phi \ket{v_k}  } { \left( \frac{NE_{jk}}{2\pi} \right)^2 - 1}   e^{-iNE_{jk} }
.  \label{eqn:V_jk}
\end{align}

To summarize, for the Hamiltonian in equation \ref{eqn:Hquantum} but with $\mu =\mu'$ 
so that it has a single parameter describing the strength of the time dependent perturbation, 
we computed the first term in the Magnus expansion for the propagator $\hat U_T$ (defined equation \ref{eqn:U_T}) in the interaction picture.   The result is given in  
equation \ref{eqn:Om1} as an operator and with matrix elements in equation \ref{eqn:mat_I}. 
This gave us an effective averaged perturbation in the Schr\"odinger representation 
with matrix elements that are given in equation \ref{eqn:V_jk}. 

\subsection{Numerical estimates of the perturbation in the interaction picture \label{sec:nummat}}

To test whether the matrix elements $V_{jk}$ (of equation \ref{eqn:V_jk})  match the behavior exhibited by 
the Floquet propagator of the periodically perturbed Harper model (equation \ref{eqn:U_T}), we display matrix elements of the Floquet propagator $\hat U_T$ along with  matrix elements of $\hat V$ in Figure \ref{fig:puf}.
Using the Trotter-Suzuki decomposition (described in appendix \ref{sec:trotter}) 
we create the operator 
\begin{align}
\hat U(\tau) = {\mathcal T} e^{-\frac{iN}{2\pi} \int_0^\tau \hat h_{\rm PH}(\tau') d \tau'} \label{eqn:Utau}
\end{align}
 integrated to times $\tau = \pi/2, \pi, 3\pi/2, 2 \pi$.  
 At the last time value, we have the full Floquet propagator $\hat U(2 \pi) = \hat U_T$ (of equation \ref{eqn:U_T}).
 For each of these operators we compute the magnitude of the matrix elements 
 $|\bra{v_j} \hat U(\tau) \ket{v_k}|$ for $j,k \in [0, N-1]$.    Here $\ket{v_j}$ are the eigenstates 
 of the unperturbed Hamiltonian $\hat h_0$. 
Each pixel color in each panel shows  the magnitude of a single matrix element, 
 with pixels in order of the eigenstate energies. 
The times $\tau$ for each operator are shown in white on the lower left in each panel. 
The parameters for the periodically perturbed Harper Hamiltonian model 
 (with Hamiltonian defined in equation \ref{eqn:Hquantum}) 
 are printed on the top of the figure.  For these models we set $\mu' = \mu$.  
We compute elements for the case $a = \epsilon$ that has a single separatrix energy 
 and a case with $a \ne \epsilon$  which has 
 two separatrices.  The rightmost panels in Figures \ref{fig:puf}a and c show horizontal and vertical 
 red lines at the separatrix energies. 
Figures \ref{fig:puf}a and c illustrate that the propagator 
 has off-diagonal elements (in the eigenbasis of the unperturbed Hamiltonian $\hat h_0$) near
 the separatrix energies. This is consistent and expected since ergodic regions in both quantum and associated classical system cover these separatrices. 
 
In Figure \ref{fig:puf}b we similarly show the magnitudes of the matrix elements $|V_{jk}| $ computed
 via equation \ref{eqn:V_jk} for the same Hamiltonian parameters as in Figure \ref{fig:puf}a.
Figure \ref{fig:puf}d is similar to Figure \ref{fig:puf}b except for the Hamiltonian parameters 
 of Figure \ref{fig:puf}c.   
Figures \ref{fig:puf}b and d show that the perturbation operator $\hat V$ also has off-diagonal  
 elements (in the $\hat h_0$ eigenbasis), near the separatrices. 
In this way the operator $\hat V$ is similar to the Floquet propagator.  
This implies that the approximations we have
 made in computing it (using only first operator in the Magnus expansion) 
have captured the key ability to cause perturbations at and near the separatrix energies.    
 
Examining equations \ref{eqn:Om1}, \ref{eqn:mat_I}, and  \ref{eqn:V_jk}, it is not obvious why 
 the eigenstates of the unperturbed Hamiltonian $\hat h_0$ near the separatrix tend to give the largest off-diagonal matrix elements for $\hat h_{F,I}$ and $\hat V$. 
\citet{Strohmer_2021} approximated the energy 
 spectrum of the unperturbed Hamiltonian $\hat h_0$ with a linear function (see Figure 
 \ref{fig:vanil}c),  however our
 numerical computations show that the spacing between energy levels is  
reduced near the separatrices.   This is expected as the phase space volume covered 
by a small range in energy $dE$ is sensitive to the geometry of the Hamiltonian level curves and 
increases near the separatrix orbit.  This is related to the relation between the derivative of 
the action with respect to energy and orbital period, with both quantities approaching infinity at the separatrix. 
As the volume in phase space sets energy level spacing via Bohr-Sommerfeld quantization, 
the increase in phase space volume within a particular $dE$,  
leads to a decrease in the energy spacing between eigenstates. 

The factors that depend upon energy differences in equations 
  \ref{eqn:mat_I} and \ref{eqn:V_jk} alone do not give larger magnitudes in $V_{jk}$ 
  for the eigenstates with energies similar to the separatrices.   
The matrix elements $\bra{v_j} \cos \hat \phi \ket{v_k}$ (in equations \ref{eqn:mat_I} and \ref{eqn:V_jk}) and the matrix elements of the commutators $[(\hat h_0)^n, \cos \hat \phi]$ (in equation \ref{eqn:Om1}) 
 are banded in the sense that they decay rapidly away from the diagonal;  
 (we found numerically that their magnitude rapidly drops as a function of increasing $|j-k|$).   
The banded behavior of the matrix with elements $\bra{v_j} \cos \hat \phi \ket{v_k}$ means that 
 the matrix elements of $\hat V$ given by 
 equation \ref{eqn:V_jk} are largest near but not on the diagonal which vanishes due to the factor
 $e^{iNE_{jk}} - 1$. 
 The denominator in equation \ref{eqn:V_jk},  $\left(\frac{N E_{jk}}{2 \pi}\right)^2 - 1$, is smallest for the terms near the separatrix where the difference between neighboring energy levels is smallest. 
Despite the simplicity of the Harper Hamiltonian when written in terms of 
clock and shift operators (see appendix \ref{sec:clock_shift}), we lack a simple analytical form 
for the separatrix eigenstates, making it difficult to see exactly why states near the separatrix 
energy tend to gives large values in equations \ref{eqn:mat_I} and \ref{eqn:V_jk}.

\begin{figure*}[htbp]
\centering
\includegraphics[width=7 truein]{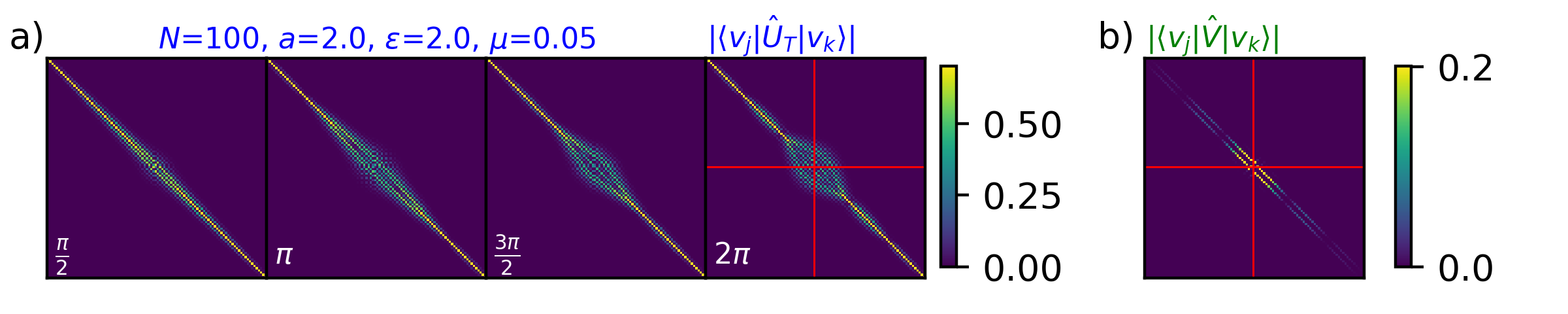}
\includegraphics[width=7 truein]{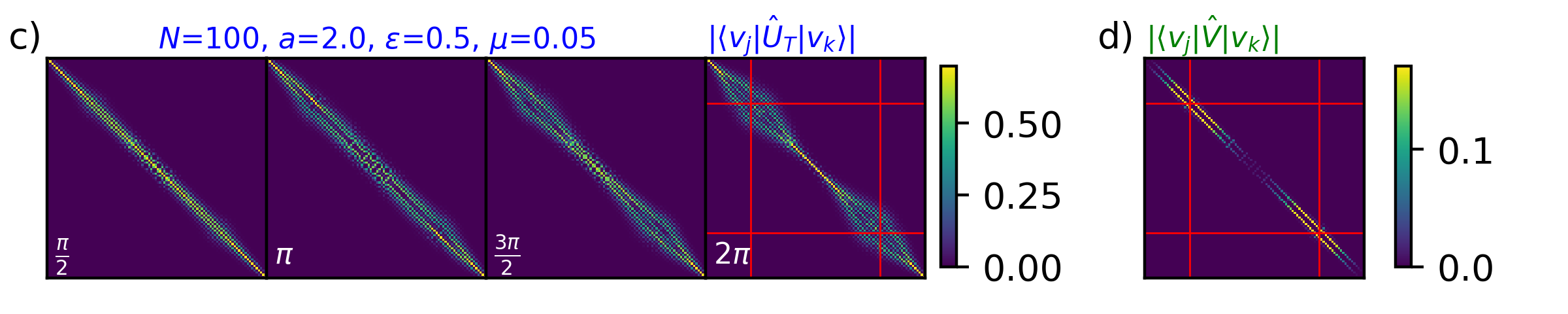}
\caption{a) The four panels on the left show the magnitudes of 
the matrix elements $|\bra{v_j}\hat U (\tau) \ket{v_k}|$ (equation \ref{eqn:Utau}) 
constructed using eigenstates of unperturbed Hamiltonian $\hat h_0$.    
Each pixel shows different values of $j, k$ 
and the eigenstates are in order of their energy. 
The 4 panels show $\hat U(\tau)$ integrated (via Trotterization) to 4 different times $\tau = 0, \pi/2, \pi, 2 \pi$
with the Floquet propagator $\hat U_T$ (integrated to the full period of $2\pi$) shown in the rightmost panel.
In that same panel we show with red vertical and horizontal lines the locations of the eigenstates that
have energies of the separatrices.   The Floquet propagator $\hat U_T$ has strong off-diagonal elements 
near these separatrices.  The parameters of the perturbed Harper Hamiltonian (equation \ref{eqn:Hquantum})  
are printed on top in blue.  For this model because $a = \epsilon$, there is a single separatrix energy. 
b) For the same Hamiltonian shown in a) we show 
 the magnitude of the matrix elements of $\hat V$ (equation \ref{eqn:V_jk}), 
also constructed from eigenstates of $\hat h_0$.  
Again the off-diagonal elements are 
strongest near the separatrices which are shown with  red horizontal and vertical lines.   
c) Similar to a) except for a Hamiltonian with different values of parameters $a, \epsilon$. 
Because $a \ne \epsilon$, there are two separatrix energies. 
d) Similar to b) except for the Hamiltonian of panel c). 
The Floquet propagator $\hat U_T$ and averaged perturbation operator $\hat V$ both contain
off diagonal elements near the separatrices, indicating that $\hat V$ 
behaves similar to the Floquet propagator and can account for ergodicity 
generated near the classical separatrix orbits. 
\label{fig:puf}}
\end{figure*}

\subsection{Relating matrix elements of the perturbation to the energy dispersion \label{sec:dispT}}

We relate the matrix elements in equation \ref{eqn:V_jk} of the perturbation $\hat V$ 
to the dispersion $\sigma_{h0,j}$ of the unperturbed 
Hamiltonian $\hat h_0$, that we numerically computed (in section \ref{sec:disp}) 
with the eigenstates of the propagator, $\hat U_T$. 
Equation \ref{eqn:hV} implies that 
the eigenstates of $\hat U_T$ are the same as the eigenstates of the operator 
\begin{align}
\hat h_0 + \mu \hat V. \label{eqn:g_F}
\end{align} 

In appendix \ref{sec:pert} we show how to compute the dispersion of an unperturbed 
energy using the eigenstates of a non-degenerate and perturbed energy operator. 
The Harper Hamiltonian $\hat h_0$ can be written as a cyclic symmetric tridiagonal matrix with non-zero off-diagonal elements (see appendix \ref{sec:clock_shift}). 
The operator is nearly tridiagonal and tridiagonal matrices 
with non-zero off-diagonal elements have distinct eigenvalues (Lemma 7.7.1 \citep{Parlett_1998}).  
The approximate eigenvalues estimated by \citet{Strohmer_2021} 
suggest that the eigenvalues are usually distinct.  Our numerical calculations confirm this expectation  
except for the case when $N$ is a multiple of 4.  
Equation \ref{eqn:g_F} is in the form of equation \ref{eqn:pert}, derived in appendix \ref{sec:pert}. 
We apply equation \ref{eqn:second} to estimate the dispersion $\sigma_{h0,j}$ (equation \ref{eqn:sigh0}) 
of $\hat h_0$  but  computed with 
the eigenstates of $\hat h_0 + \mu \hat V$  (which are also the eigenstates of $\hat U_T$) 
\begin{align}
\sigma_{h0,j}^2 
& = \mu^2
 \sum_{j\ne k} |   V_{jk}  |^2 
\label{eqn:sigh02}
.
\end{align}

\begin{figure}[htbp]
\centering
\includegraphics[width=3.3 truein, trim = 0 0 0 0, clip]{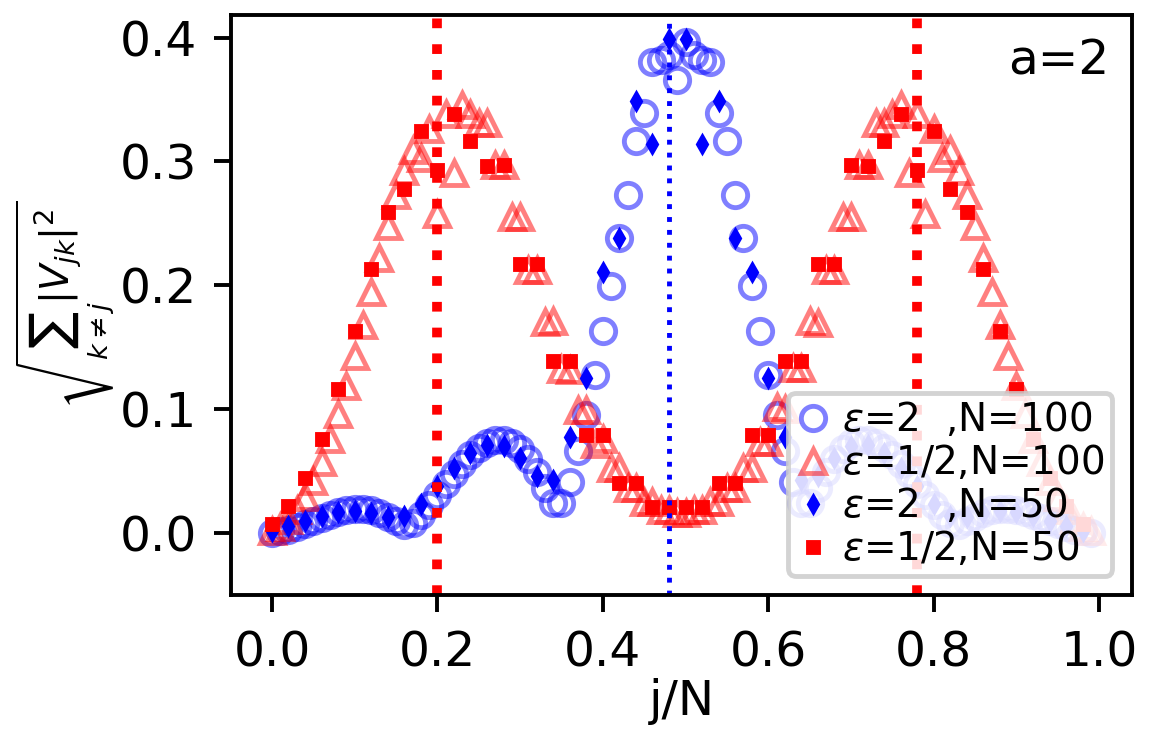}
\caption{We show estimates for 
the energy standard deviation $\sigma_{h0,j}/\mu$ for 4 different quantum systems 
(with Hamiltonian in the form of equation \ref{eqn:Hquantum}) and as a function of index $j/N$.   The standard deviations are computed using equations \ref{eqn:V_jk}  and \ref{eqn:sigh02}  using the eigenstates of the unperturbed Hamiltonian $\hat h_0$. 
Parameters for the Hamiltonian are shown in the key and on the upper right. 
As in Figure \ref{fig:puf}, we compute the energy dispersion for models with two separatrices
(open red triangles and small red squares) and models with a single separatrix (with $a=\epsilon=2$ and shown with open blue circles and filled small blue diamonds).  The open points show models with 
number of states $N=100$ and the filled points show models with $N=50$.  The energy standard deviation is 
not strongly dependent upon the number of states.   States with energy near the separatrices are shown 
with vertical dotted lines.   The energy standard deviations peak near states 
that have the energy of the separatrix in the associated classical model.    The energy dispersion 
$\sigma_{h0,j}$ provides an estimate for the ergodicity of the eigenstates of the propagator $\hat U_T$ 
(as discussed in section \ref{sec:disp})
and can be estimated from the matrix elements of the perturbation in the interaction picture 
(as discussed in section \ref{sec:ave}).
 \label{fig:sqrtVjk}
 }
\end{figure}

We numerically compute $\sigma_{h0,j}/\mu$ using the matrix elements given in equation 
\ref{eqn:V_jk}.  The results are plotted as a function of $j/N$ for 4 different unperturbed Hamiltonians
in Figure \ref{fig:sqrtVjk}.
As we saw in Figure \ref{fig:puf}, peaks in Figure \ref{fig:sqrtVjk} also occur at the separatrices. 
These correspond to estimates 
 for the standard deviation of the unperturbed energy operator computed from the 
 eigenstates of the propagator, hence they are equivalent to those we computed 
 from the numerically computed propagators and discussed in section \ref{sec:disp}. 
The peak values for $\sigma_{h0,j}/\mu$ we computed are fairly large, about 0.4. 
This size is large enough to 
be approximately consistent with the size of the standard deviations we measured directly from the eigenstates of the Floquet propagator in similar models  
(those shown in Figure \ref{fig:Q9b}c, and \ref{fig:Q9c}c).  

Using equation \ref{eqn:V_jk} 
\begin{align}
\sigma_{h0,j}^2  
& =\mu^2 \sum_{k\ne j} \left(  \frac{N E_{jk}}{4 \pi^2} \right)^2 
\frac{ |\bra{v_j} \cos \hat \phi \ket{v_k} |^2}{ \left( \left(\frac{NE_{jk}}{2 \pi} \right)^2 - 1 \right)^2}
2( 1 - \cos (N E_{jk})). \label{eqn:gen}
\end{align}
As discussed at the end of section \ref{sec:nummat}, we found numerically that the 
sum is dominated by terms with $k$ near but not equal to $j$
because  $|\bra{v_j} \cos \hat \phi \ket{v_k} |$ decays away from its diagonal which vanishes
due to the cosine factor.  We suspect that the separatrix dominates 
 because the spectrum of $\hat h_0$  
has smaller energy level differences near the separatrix and this causes a small denominator
 in equation \ref{eqn:gen}. 
 
The adiabatic or low perturbation frequency limit is defined with the ratio of perturbation frequency 
to pendulum libration frequency  $\lambda = \nu/\omega_0$, which for our dimensionless Harper model
is equal to $ \omega_0^{-1}$ because we work with time in units of inverse perturbation frequency. 
In the non-adiabatic (or rapid perturbation) limit, classical estimates for the 
half width of the chaotic region decay exponentially 
(the $\omega_0 $ small limit in equation \ref{eqn:DeltaH0}) 
with chaotic region width $\Delta H \propto e^{- \frac{\lambda\pi}{2}}$ \citep{Chirikov_1979}. 
If we adopt a characteristic energy $\tilde \hbar \omega_0 = \frac{2\pi}{N} \omega_0$ between 
states, then $\epsilon_{jk} = \frac{N E_{jk} }{2 \pi} \sim  \omega_0 (j-k)$.  
In the limit of small $\omega_0$, equation \ref{eqn:gen} becomes 
$\Delta H \sim \epsilon_{jk}^4 \sim \omega_0^4$.  This decays rapidly as $\omega_0 \to 0$, 
consistent with the classically derived estimate, but with a 
 steep power-law form rather than an exponentially decaying form.   
In the adiabatic limit, ($\omega_0$ large), 
equation \ref{eqn:gen} gives $\Delta H \propto \epsilon_{jk}^{-2} \propto \omega_0^{-2}$
and is similar to the classical expression, (though see \citep{Shevchenko_2008,Soskin_2008,Soskin_2009,Shevchenko_2012} for improved 
calculations in the adiabatic limit). 


Based on the estimates in this subsection and subsection \ref{sec:ave}, we summarize 
how we estimated the energy width of the ergodic region for a Floquet quantum system. 
We consider a Hamiltonian $\hat  h_0 + \hat h_1(t)$ 
with perturbation $\hat h_1(t)$ that is periodic with period $T$.  
We identify the energy $E_s$ of a separatrix orbit 
in an associated classical system (that contains a hyperbolic fixed point) 
and an eigenstate $\ket{v_s}$ of the unperturbed system $\hat h_0$ that has energy close to that
 of the separatrix orbit;  $\bra{v_s} \hat h_0 \ket{v_s} \approx E_s$.
We use the first term in the Magnus expansion within the interaction picture 
to compute an averaged perturbation (equations \ref{eqn:h_1I}, \ref{eqn:UTI}, \ref{eqn:h_F}, \ref{eqn:hFI_def}, \ref{eqn:Om1_a}, and \ref{eqn:FF}). 
Transfering back into the Schr\"odinger picture, we estimate the width of the ergodic region near 
the separatrix using equation \ref{eqn:sigh02} and giving 
a width in energy 
\begin{align}
\Delta H \sim 
\sqrt{ \sum_{k\ne s} 
\left|\bra{v_s}\frac{1}{T} \int_0^T dt e^{ \frac{i}{\hbar} \hat h_0 t} \hat h_1(t) e^{- \frac{i}{\hbar} \hat h_0 t}  \ket{v_k}\right|^2}  \label{eqn:DeltaH_Q}
\end{align}
where $\ket{v_k}$ are the eigenstates of $\hat h_0$. 
The sum is likely to be dominated by matrix elements for states $\ket{v_k}$ that have 
energy close to $E_s$, that of the separatrix. 

Equation \ref{eqn:DeltaH_Q} for estimating the width (in energy) of a chaotic region 
at a separatrix is analogous to equation \ref{eqn:DeltaH_C} in the classical setting 
(following \citep{Melnikov_1963,Zaslavsky_1968,Chirikov_1979,Shevchenko_2000,Treschev_2010}).  
Instead of using the separatrix orbit of the unperturbed classical Hamiltonian, 
equation \ref{eqn:DeltaH_Q} uses an eigenstate 
of the unperturbed system that is near the energy of the classical separatrix orbit. 
Instead of integrating the perturbation along the separatrix orbit, equation \ref{eqn:DeltaH_Q} 
sums over states with energies near that of the separatrix and the perturbation 
is averaged while undergoing evolution from the unperturbed system. 

\section{Summary and Discussion}

We have explored a doubly periodic in space (in angle and momentum) and periodic in time Hamiltonian dynamical system that can exhibit a hybrid or mixed phase space with both regular
(non-ergodic) and chaotic (ergodic) motion in both classical and quantum settings.  
Our system is a periodically perturbed Harper model and is closely related to the periodically perturbed
pendulum model discussed by \citet{Chirikov_1979}. 
We chose to work with a compact
phase space so that we can work in finite dimensional quantum spaces, 
  facilitating numerical computation of the quantum Floquet progagator.  
 We quantize the classical model by converting angle and momentum to a pair of operators 
 that are related via discrete Fourier transform and using Bohr-Sommerfeld quantization to set 
 the number of states from the area in phase space and Planck's constant.  
  If the dimension of the system is odd, then this  quantization method is equivalent 
 to Wigner-Weyl quantization via the same point operator we use to make 
 discrete versions of coherent states. 

We used a discrete version of coherent states to compute Husimi functions showing phase space 
for the eigenstates of the Floquet propagator derived from the periodically perturbed Harper Hamiltonian.  
A comparison between Husimi functions, and surfaces of section created 
for the associated classical system by directly integrating orbits, shows remarkable similarity between 
morphology of orbits and eigenstates in phase space. 
We find that the expectation value for an eigenstate of the unperturbed energy operator 
and the mean value for an orbit of the unperturbed
energy function can be used to match eigenstates of the Floquet propagator to similar classical orbits. 
Moreover the energy dispersions computed for the Floquet propagator's eigenstates are similar to those computed 
from the classical orbits.  The energy dispersion can be used to estimate which  
orbits and eigenstates are ergodic and gives an estimate for the width of a chaotic region that 
is generated near a separatrix of the unperturbed system. 

We derived an estimate in the quantum system for the width of the ergodic region (equation \ref{eqn:DeltaH_Q}) that is analogous to that derived in the classical setting where the perturbation is integrated along the separatrix orbit.   
Like the classical 
expression, our expression is first order in the perturbation strength.  Instead of integrating along
the separatrix orbit, our expression involves an average of the perturbation in the interaction representation,
is derived from the first term in the Magnus
expansion, and is evaluated using eigenstates with energy near that of the separatrix. 
The width of the separatrix chaotic layer is only roughly approximated  
by the Melnikov-Arnold integral and recent studies have improved upon this estimate  (e.g., \citet{Soskin_2009}).    Future studies 
could perhaps improve upon our estimate in the quantum setting. 

Time dependent perturbation theory in quantum mechanics can fail to capture the behavior
of a system because terms involved in the expansions are not small.  We attempted 
Dyson series and Magnus expansions in the Schr\"odinger picture but they failed to give a model 
that captured the behavior near the separatrices seen in the periodically perturbed Harper model. 
The first term in the Dyson series is first order in the perturbation strength but the resulting computed operator
is not unitary.  Nonzero and complex high index terms in the Magnus expansion in the Schr\"odinger picture contain commutators that are first order in the perturbation strength. 
However, we found that the first term in the Magnus expansion in the interaction picture 
is first order in the perturbation strength and gives behavior 
similar to that seen in our system with large off-diagonal elements (in the eigenbasis of
the unperturbed Hamiltonian) in the propagator at energy levels near that of the separatrices. 
Unfortunately, this Magnus expansion is only guaranteed to converge at a perturbation 
parameter below those used in our numerical exploration.   A number 
of studies have found that the first few terms in a Magnus expansions 
are often useful even the full expansion does not converge (e.g., \citep{Blanes_2009,Kuwahara_2016}).  
Perhaps the averaged perturbation in the interaction picture is 
large in the same region where the propagator is large within the stationary phase approximation
or the result is close to the correct one in a geometric sense. 
The stationary phase approximation  
has been effective for derivation, in the semi-classical limit,  of the van Vleck propagator \citep{VanVleck_1928}.  
A geometric approach (e.g., \citet{Paivi_2023,Schindler_2025}) could help with improved 
perturbative methods for 
chaotic Floquet systems, like the one we studied here.   Time-periodic classical Hamiltonian 
systems can be embedded in an extended phase space where they appear time-independent. 
Methods that use an extended Floquet Hilbert space (e.g., \citet{RodriguezVega_2018}) which is 
akin to the $(t,t')$ formalism (e.g., \citep{Grifoni_1998}) could be used to derive a time-independent perturbation theory.  
These and other approaches could be explored in future work on Floquet systems such as the perturbed Harper model.  

The Hamiltonian of the quantum Harper model is remarkably simple in terms of 
clock and shift operators (see appendix \ref{sec:clock_shift}).  Its eigenstates can 
be approximated with Hermite polynomials \citep{Strohmer_2021}. Nevertheless,   
 we had difficulty analytically calculating a quantum counterpart 
to the Melnikov-Arnold integral which gives an estimate for the width of the chaotic 
region.  With more insight perhaps it could be possible 
derive a simpler analytical estimate  
for the ergodic region width leveraging our approach (in section \ref{sec:ave} and \ref{sec:dispT}).  
Other Hamiltonian systems, such as that studied by \citet{Yampolsky_2022}, have simpler expressions for their eigenstates and for these it might be easier to derive analytical estimates for the width of ergodic regions. 
 
The resonance overlap criterion \citep{Chirikov_1979} for the onset of classical 
chaos has recently been generalized to a quantum system \citep{Yampolsky_2022}.  
The system of hard core particles in a box, studied by 
\citet{Yampolsky_2022} displayed 
contiguous patches of nearly unperturbed eigenstates, separated by strongly perturbed states. 
The periodically perturbed Harper model, studied here, exhibits similar phenomena,  in that it also displays perturbed eigenstates, that have broad and diffuse ergodic Husimi distributions, 
separated by states that have narrow Husimi distributions that are similar to regular classical orbits. 
\citet{Yampolsky_2022} used matrix elements of the perturbation, computed using eigenstates 
of the unperturbed system (their equation 5) to estimate the width 
of the chaotic layer near resonance.  Our expressions, such as 
equation \ref{eqn:DeltaH_Q} similarly includes an expectation of the 
perturbation at the separatrix to estimate the width of a chaotic region. 
In our study we did not focus on the resonance overlap criterion for the onset of
chaos, rather we focused on the role of 
chaos induced at a separatrix, which was part of 
the derivation of the resonance overlap criterion \citep{Chirikov_1979}. 
Our study of the periodically perturbed Harper model 
 is complimentary as we focused on discrete and time-periodic systems
potentially more directly leading to applications 
in the realm of quantum computing and Floquet engineering. 

Chirikov's resonance overlap criterion and estimates for the Lyapunov time based on the periodically 
perturbed pendulum model have been widely applied to predict stability and time-scales for evolution in celestial mechanics 
(e.g., \citep{Wisdom_1984,Mardling_2008,Quillen_2011,Hadden_2018,Petit_2020}). 
The strong connection between the behavior of the classical and quantum 
Harper model suggests that classical methods  
for predicting chaotic phenomena might inspire methods that can be applied to some quantum systems.  


\vskip 1 truein

{\bf Acknowledgements}

We thank Alex Iosevich, Esteban Wright, Damian Sowinski,  Nathan Skerrett and Machiel Blok for helpful and interesting discussions. 
We thank Taco Visser and Liz Champion for introducing us to coherent states. 
We thank Ray Parker for insightful discussions on the interaction picture and the Magnus expansion. 
We thank Joey Smiga for discussions on quantum ergodicity and helping us improve this manuscript. 

{\bf Conflict of interest statement}

The authors have no conflicts to disclose.

{\bf Author Contributions:}
The project was conceived and carried out by A. Quillen. 
A. Miakhel contributed to numerical computations of the quasi-energy distributions and
with analytical calculations of the separatrix width. 

{\bf Data Availability Statement:}

The python jupyter notebooks written and used in this study are openly available on the public repository  \url{https://github.com/aquillen/Qperio}

\bibliographystyle{elsarticle-harv}
\bibliography{Qchaos}

\appendix

\section{Discrete operators \label{sec:discrete}}

The position operator $\hat \phi$ is a diagonal operator in the $\{ \ket{j} \}$ basis 
and the momentum operator $\hat p$ is a diagonal operator in the Fourier basis;  $\{ \ket{m}_F \}$. 
In analogy to a continuous system, we use two discrete operators, one corresponding to
a translation in momentum, and the other giving a translation in real space,   to 
generate a displacement operator. In analogy with coherent states for a continuous system,  
discrete coherent states are created via operating with the displacement operator on a minimum 
uncertainty state that resembles a Gaussian. 

The unitary $\hat Z, \hat X$ operators, introduced by Schwinger in his work on mutually unbiased bases \citep{Schwinger_1960},  
(also see \citet{Bjork_2008})
\begin{align} 
\hat Z & = \sum_{n=0}^{N-1} \omega^n \ket{n}\bra{n} = \sum_{k=0}^{N-1} \ket{k+1}_F\bra{k}_F \\
\hat X &=  \sum_{k=0}^{N-1} \omega^{-k}  \ket{k}_F\bra{k}_F = \sum_{n=0}^{N-1} \ket{n+1}\bra{n} 
\label{eqn:XZdefs}.
\end{align}
In these expressions, addition for the state vectors is modulo $N$. 
These are equivalent to a set of generalized Pauli matrices called {\it clock and shift operators} or Weyl-Heisenberg matrices \citep{Appleby_2005}. 
These describe shifts in a periodic discrete quantum space. 
These operators have been used to study the quantized
Baker map on the torus, and in phase space \cite{Balazs_1989} and in quantum information theory (e.g.,  \citet{Bjork_2008}). 

\section{Husimi distributions constructed from discrete coherent state analogs \label{sec:Hus}}

To compare the quantum and classical systems, we compute Husimi distributions (also called
the Husimi Q representation) which are quantum analogs to classical phase space distributions \citep{Cartwright_1976}. 
Our procedure for constructing Husimi distributions on a torus (a doubly periodic space)
uses a discrete displacement operator, following  \cite{Balazs_1989,Saraceno_1990,Galetti_1996,Rivas_1999,Rivas_2002,Bjork_2008}.   
The displacement operator is used 
to construct discrete analogs for coherent states which in turn are used to compute  Husimi distributions.  

A displacement operator is 
 \begin{align}
 \hat D(k,l) = \omega^{-kl/2} \hat Z^k \hat X^l 
 \label{eqn:Dkl}.
 \end{align}
 for integers $k,l \in \{0, 1, .....,  N-1 \} $. 
Because the operator $\hat X$ (described in appendix \ref{sec:discrete}) acts like a raising operator in the $\ket{n}$ basis 
 and $\hat Z$ acts like a raising operator in the $\ket{k}_F$ basis, the displacement operator $D(k,l)$ 
translates in position and momentum. 
 
Following \citet{Galetti_1996}, to generate coherent state analogs, we start with 
 a unimodal eigenstate of the discrete Fourier transform 
\begin{align}
\ket{\tilde \eta} & = b_\eta \sum_{m=0}^{N-1}  \vartheta_3 \left( \frac{\pi m}{N}, e^{-\pi/N} \right)  
\ket{ m }\nonumber  \\  
& \propto \sum_{m=0}^{N-1} \sum_{j=-\infty}^\infty e^{-\frac{\pi}{N} (m + jN)^2} \ket{m} .  \label{eqn:eta}
\end{align}
Here  $\vartheta_3$ is a Jacobi theta function and the coefficient $b_\eta \approx \left(\frac{2}{N} \right)^\frac{1}{4}$  serves to normalize the state vector. 
In the limit of large $N$, this becomes a Gaussian function which is the state vector with the minimum 
possible uncertainty in phase space. 
 
Because the state $ \ket{\tilde \eta}$ is equal to its Fourier transform and resembles a Gaussian,  
it acts like a state of optimal uncertainty (like a Gaussian function in the continuous setting).   
The expectation value 
$$\bra{\tilde \eta} f(\hat X) \ket{\tilde \eta} = \bra{\tilde \eta} f(\hat Z) \ket{\tilde \eta}$$
for any polynomial function $f()$, 
because $\hat Z = \hat Q_{FT} \hat X \hat Q_{FT}^\dagger$ and $\hat Q_{FT} \ket{\tilde \eta} = \ket{\tilde \eta}$. 
So $\expval{\hat X^2}  = \expval{\hat Z^2}$ for the state $\ket{\tilde \eta}$. 

We compute discrete analogs for coherent states using the displacement operator 
of equation \ref{eqn:Dkl} and the state $\ket{ \tilde\eta}$ (defined in equation \ref{eqn:eta})
\begin{align}
 \ket { k,l} = {\hat D} (k,l) \ket{ \tilde \eta}  \label{eqn:coherent}
 \end{align}
 for each $k,l \in \{ 0, .... , N-1 \} $.  
We compute 
the Husimi function of a state vector $\ket{\psi } $ 
 \begin{align}
H_Q(k,l) = |\braket{\psi}{ k,l} |^2.  \label{eqn:Hus}
\end{align} 
By computing $H_Q$ for $k,l \in \{ 0, 1, ....,  N-1 \} $ we produce 
 an $N\times N$ grid of positive values representing the phase space distribution 
of the state $\ket{\psi}$. 

\section{Wigner-Weyl quantization \label{sec:WigWeyl}} 

The way we quantized the perturbed Harper model,  
taking $H \to \hat h$ described at the beginning of section \ref{sec:qm}, is equivalent to quantization via
a discrete Weyl symbol or (also called a discrete Wigner-Weyl transform), (e.g., \citet{Wootters_1987}).  
Following \citet{Bjork_2008},  a point operator $\hat A_{nk} $ for a discrete system can be constructed 
from the displacement operator that resembles the Wigner-Weyl transform of a continuous system; 
\begin{align}
 \hat A_{nk} & = \frac{1}{N} \sum_{x=0}^{N-1} \ket{n+x} \bra{n-x} \omega^{2xk} \\
 & = \frac{1}{N} \hat X^n  \hat Z^{k} \hat P \hat Z^{-k} \hat X^{-n} \nonumber \\ 
 & =\frac{1}{N}   \hat D(k,n) \hat P \hat D(k,n)^\dagger.   \label{eqn:Ank}
\end{align}
for integers $n,k \in \{0, 1, .... N-1 \}$ and 
 with parity operator 
 \begin{align}
 \hat P = \sum_{x=0}^{N-1} \ket{x}\bra{-x\!\! \mod N}.  \label{eqn:parity}
 \end{align}
The operators $\hat X, \hat Z$ are described in appendix \ref{sec:discrete}. 
The Weyl transform of a function of discrete phase space denoted with integers $a(k,n)$ gives an operator 
\begin{align}
\hat a = \sum_{k,n = 0}^{N-1} a(k,n) \hat A_{nk} .  \label{eqn:Wig_transform}
\end{align}
 Only if 
 $N$ is odd does $\hat A_{nk}$ gives a basis for operators w.r.t to the Frobenius or Hilbert-Schmidt norm;  
 \begin{align}
 \tr (A_{nk}^\dagger A_{n'k'}) = \delta_{n,n'} \delta_{k,k'} \  {\rm iff \ } N \ {\rm odd} .
 \end{align}
The following sums are valid for $N$ odd; 
\begin{align}
\sum_{k=0}^{N-1} \hat A_{nk}  &= \sum_{x=0}^{N-1} \ket{n+x} \bra{n-x} \sum_{k=0}^{N-1} \omega^{2 kx} \frac{1}{N}  =  \ket{n}\bra{n}  \nonumber \\
\sum_{n=0}^{N-1} \hat A_{nk}  &=  \ket{k}_F\bra{k}_F \label{eqn:Asums}
\end{align}
so the first index of the point operator $\hat A_{nk}$  is a coordinate index and the second index is a momentum index. 

Using $v(n,k) = V(n)$ corresponding to a potential energy function, we use equations \ref{eqn:Ank} - \ref{eqn:Asums}
to create the  operator $\hat v = \sum_n V(n) \ket{n}\bra{n}$ using the Wigner-Weyl transform.  
With $t(n,k)  = T(k)$ corresponding to a potential energy function in the discrete space, we find that $\hat t =  \sum_k T(k) \ket{k}_F\bra{k}_F$ 
in the Fourier basis.  Thus Weyl quantization of the separable Hamiltonian such as that in equation \ref{eqn:Hclassical} 
with $T(k) = a ( 1 - \cos( 2 \pi k/N)) $ gives the same Hamiltonian operator as in equation \ref{eqn:Hquantum}, but only for odd $N$ discrete dimension.   \citet{Leonhardt_1996} showed how
to create a well-behaved discrete point operator (with $2N \times 2N$ points)  if the dimension $N$ is even. 

Quantization of a separable Hamiltonian using a pair of mutually unbiased bases that are related via
discrete Fourier transform, 
discussed in section  \ref{sec:qm}, and the discrete coherent states for constructing Husimi distributions, 
(discussed in appendix \ref{sec:Hus}) can be done for dimension $N$ either even or odd.  
However Weyl quantization via the same point operator used to make coherent states 
is only equivalent to quantization via levering a pair of mutually unbiased bases if dimension $N$ is odd. 


\section{Computing the propagator via Suzuki-Trotter decomposition \label{sec:trotter}}

We numerically compute 
the unitary transformation of the Floquet propagator $\hat U_T$ in equation \ref{eqn:U_T} via Suzuki-Trotter decomposition (e.g., \citet{Hatano_2005}).
The Hamiltonian is divided into two pieces, one that contains the momentum operator and
the other that contains the angle operator, 
\begin{align}
\hat h(\tau)_{\rm PH} & =  \hat A_p + \hat B_\phi(\tau ) 
\end{align}
with 
\begin{align}
\hat A_p  & = a(1 - \cos \hat p)  \nonumber \\
& = \sum_{m=0}^{N-1} a\left( 1 - \cos \frac{2 \pi m}{N} \right)
\ket{m}_F \bra{m}_F  \label{eqn:A} \\
\hat B_\phi( \tau) & = -\epsilon \cos \hat \phi - \mu \cos ( \hat \phi - \tau) -  \mu' \cos ( \hat \phi + \tau) \nonumber  \\
& = \sum_{j=0}^{N-1}\Bigg[ -\epsilon \cos \frac{2\pi j}{N}  - \mu \cos \left(\frac{2\pi j}{N}  - \tau\right)\nonumber \\
& \qquad 
	- \mu' \cos \left(\frac{2\pi j}{N}  + \tau\right)
	 \Bigg] \ket{j}\bra{j} .\label{eqn:B}
\end{align}
We approximate unitary evolution of a duration $d \tau$ with the approximation for an 
exponential of two operators  $\hat A,\hat B$ with a symmetric second order decomposition 
\begin{align}
 e^{ (\hat A + \hat B)d\tau }  \approx e^\frac{\hat Ad\tau}{2} e^{\hat B d\tau}  e^\frac{\hat Ad\tau}{2} + {\cal O}(d\tau^3) + ... .
\end{align}

We approximate the propagator in equation \ref{eqn:U_T}, with  
evolution from $\tau = 0$ to $2 \pi$ (a period), using $n_\tau$ steps of duration $d\tau = 2 \pi/n_\tau$  
\begin{align}
\hat U_T \approx \Lambda_{A}^\frac{1}{2} \left( {\mathcal T} \prod_{j=0}^{n_\tau-1} \Lambda_{B,j}\Lambda_{A}  \right)
\Lambda_{A}^{-\frac{1}{2}}
 \end{align}
with operators 
\begin{align}
\Lambda_{A} & 
	= e^{- \frac{i N  }{n_\tau }\hat A_p }  
	\nonumber\\
\Lambda_{B,j} & 
	= e^{- \frac{i N  }{ n_\tau }\hat B_p(\frac{2 \pi j}{n_\tau} ) } 
\end{align}
with $\hat A_p, \hat B_\phi $ from equations \ref{eqn:A} and \ref{eqn:B}.
Using this approximation, we numerically compute propagators for systems with different 
parameters and in different dimensions $N$.    Because of the factor of $N$ in the argument of
of $\hat U_T$, we use $n_\tau = 5N$ Trotterization steps 
though we did not see significant differences between the properties of 
propagators computed with a few times fewer or greater numbers of steps. 
After we have numerically computed a propagator matrix,
we compute its associated eigenvectors and eigenvalues  with standard 
linear algebra computational algorithms. 

\section{Surfaces of section and Husimi functions at intermediate times \label{sec:shift} }

\begin{figure*}[htbp]\centering
\includegraphics[width=5.1 truein, trim=0 0 0 0,clip]{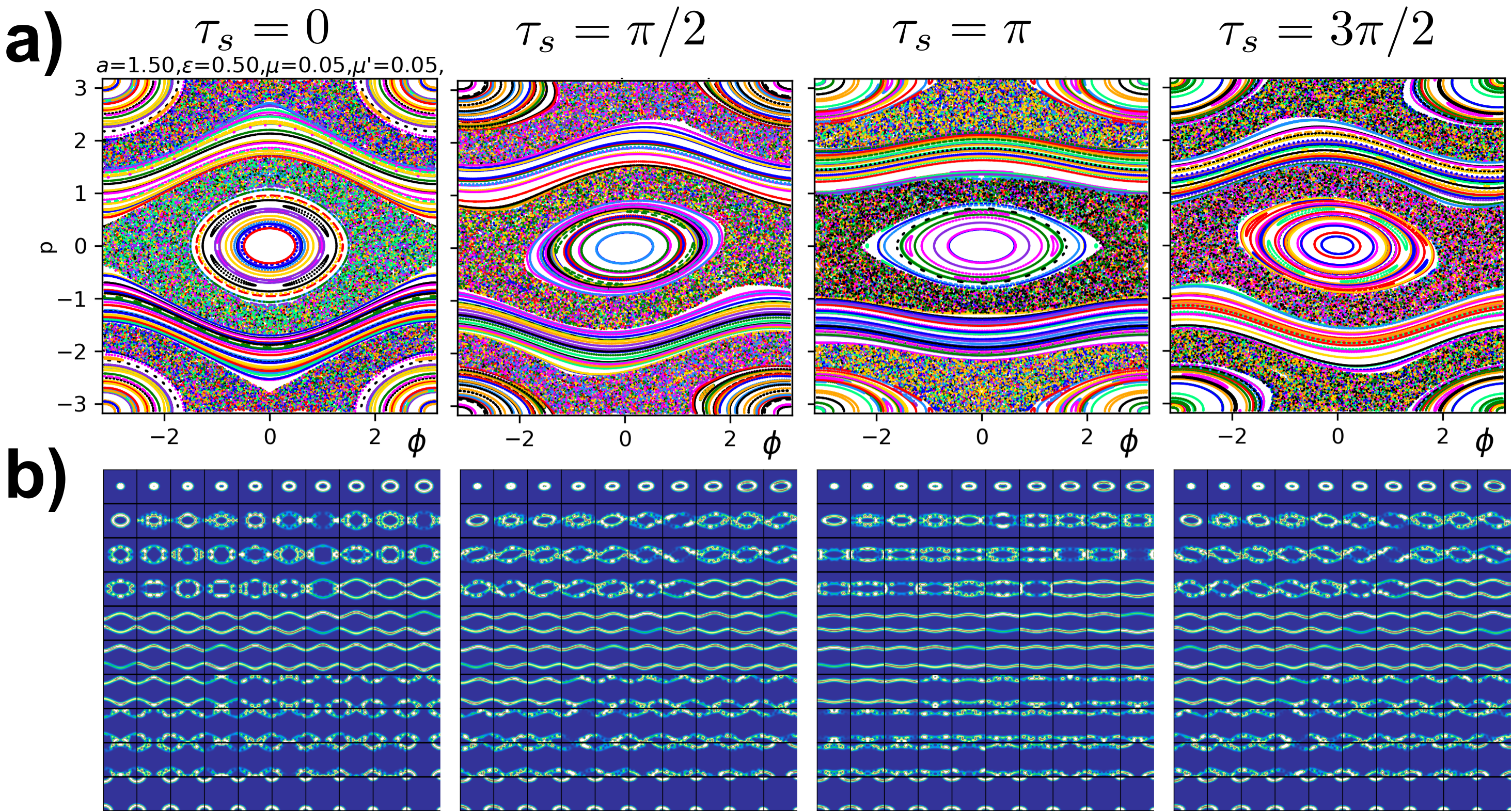}
\caption{a) We show Poincar\'e surfaces of section. 
Each panel is similar to Figure \ref{fig:Q9b}a except the surfaces
of section are generated using points at times $\tau = \tau_s $ modulo $2\pi$.  
The value of  $\tau_s$ is $0, \pi/2,  \pi$ and $3\pi/2$, for each panel, respectively, 
as shown on the top of each panel. 
 b) Below each surface of section we show 
 the Hussimi distributions of the Floquet propagator eigenstates for the associated 
quantum model with $N=100$. 
\label{fig:Q11_tt}}
\end{figure*}

The surface of section (or Poincar\'e map) shown 
in  \ref{fig:Q9b}a (and other figures) shows orbits in phase space plotted at times 
that are multiplies of $2 \pi$. In figure \ref{fig:Q11_tt}a we show Poincar\'e 
surfaces of section for the same classical Hamiltonian as shown \ref{fig:Q9b},  
however the points are at times 
$\tau = \tau_s $ modulo $2\pi$ with $\tau_s = 0, \pi/2, \pi , 3\pi/2$, for each panel, respectively. 
The classical Hamiltonian is that of equation \ref{eqn:Hquantum},  with $a =1.5 , \epsilon = 0.5$
and $\mu = \mu' = 0.05$ (and the same as shown in Figure \ref{fig:Q9b}).  
The propagator for the quantum model is shifted in time by modifying equation \ref{eqn:U_T};  
\begin{align}
\hat U_T(\tau_s)  & = {\mathcal T} e^{- \frac{i}{\tilde h} \int_{\tau_s}^{2 \pi + \tau_s} \hat h(\tau)_{\rm PH} d \tau } 
 \label{eqn:U_T_shift} .
\end{align} 
For the related quantum model with dimension $N=100$, in Figure \ref{fig:Q11_tt}b  
we show Husimi distributions, computed from the eigenvectors of the propagator   
also shifted in time. 
The morphology of the surfaces of section and Husimi distributions vary within the period.  
At intermediate times within the period, the classical orbits resemble the Husimi distributions.


\section{Quasi-energy spacing distributions \label{sec:quasi}}

\begin{figure*}[htbp]\centering
\includegraphics[width=3.4 truein, trim=0 0 0 0,clip]{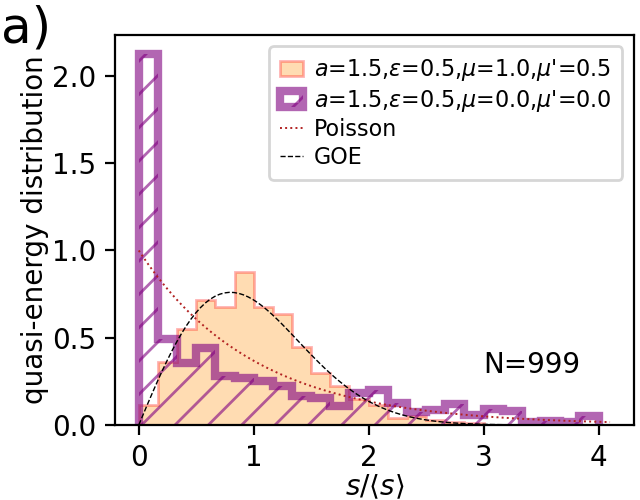}
\includegraphics[width=3.4 truein ,trim = 0 0 0 0,clip]{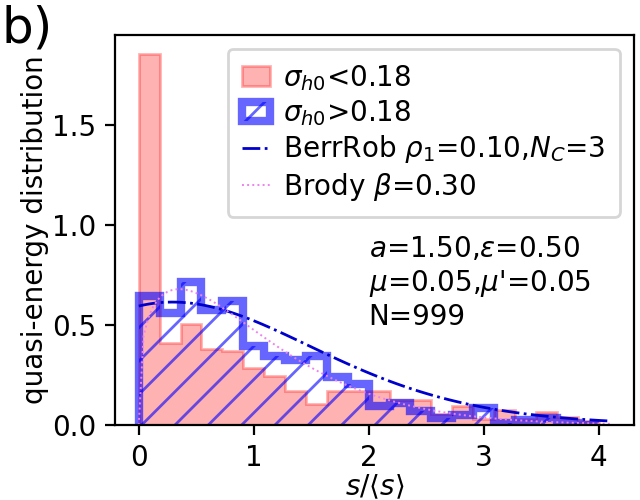}
\caption{Comparing distributions of quasi-energy spacings. 
a) On the left we compare the distributions of quasi-energy spacings for two 
quantum models.  The solid orange histogram is derived from 
a classical integral model that is fully chaotic. 
The hatched purple histogram is derived from a  classical model exhibiting chaotic regions  and is 
shown in Figure \ref{fig:Q9c}.  Histograms are normalized by the mean quasi-energy difference $\langle s \rangle$. The dotted magenta line shows a Poisson process with energy level differences $p(s) = e^{-s}$ with a mean of 1.  The dashed black line shows the spacing distribution for the Gaussian orthogonal ensemble random matrix model  (equation \ref{eqn:GOE}).  
b) We show quasi-energy spacings extracted from two subspaces in the system shown in Figure \ref{fig:Q9b}
but with $N=999$ states.   The solid pink distribution  is constructed from the subspace of eigenstates 
with a low energy dispersion,  $\sigma_{h0,j} < 0.18$, (as computed with equation \ref{eqn:sigh0}) 
and the blue hatched histogram shows the distribution of the remaining eigenstates. 
The blue dot-dashed curve shows a Berry-Robnik distribution  and the violet dotted curve shows
a Brody distribution. 
\label{fig:2hists}}
\end{figure*}

Distributions of differences $s_j = \lambda_{j+1} - \lambda_j$ between consecutive quasi-energies, normalized by their mean value $\langle s\rangle$,  are shown in Figure \ref{fig:2hists}. 
In Figure \ref{fig:2hists}a we show the quasi-energy spacing distributions for the classically regular (non-chaotic) model shown in Figure \ref{fig:vanil} as an orange filled histogram and that of the 
perturbed and similar chaotic system (with large values of $\mu, \mu'$) as a purple hatched histogram. 
The surface of section for this model is almost completely ergodic. 
In Figure \ref{fig:2hists}a, a magenta dotted line shows an exponential distribution, expected for
a Poisson process, also normalized with a mean of 1. 
The dashed black line shows the Wigner-Dyson distribution 
 \begin{align}
 p_{\rm GOE}(s) = \frac{\pi}{2} s e^{- \frac{4}{\pi} s^2} \label{eqn:GOE}
 \end{align}
 which is that of level spacings for the Gaussian orthogonal ensemble (GOE) random matrix model, 
 also normalized so that its mean is 1.  

Figure \ref{fig:2hists}a shows that the distribution of energy level spacings is sensitive to 
 the presence of chaotic regions in the model, with the regular model exhibiting a distribution 
 more similar to an exponential.  The distribution for the system containing chaotic regions  is broader
 and closer to that of a random matrix model. 
 
Figures \ref{fig:Q9b} and \ref{fig:Q9b_large} show a classical system that has two chaotic regions. 
We divide the eigenstates into two sets, 
depending upon whether their standard deviation $\sigma_{h0,j}$ is above or below 0.18.    
Using these two subspaces, we separately compute two quasi-energy spacing distributions, 
and these are shown in Figure \ref{fig:2hists}b.
The distribution for eigenstates with $\sigma_{h0,j} < 0.18$ and associated with integrable regions 
in phase space in the associated classical model, is plotted as a filled pink histogram.
The distribution for eigenstates with $\sigma_{h0,j} > 0.18$ and associated with area filling 
and chaotic orbits in the associated classical model, is plotted as a hatched blue histogram.
Figure \ref{fig:2hists}b illustrates that there is a difference between the quasi-energy distributions 
in the two subspaces.   The distribution of quasi-energy spacings is closer to that of a random matrix theory 
in the ergodic subspace and closer to a Poisson model in the regular subspace. 

The quasi-energy level spacing distribution shown in Figure \ref{fig:2hists}b is shallower 
than a Poisson distribution but is not as broad as a Wigner-Dyson distribution. 
There are a number of proposed ways to smoothly interpolate between statistical models of energy level spacings for mixed systems \citep{Brody_1981,Berry_1984,Zakrzewski_2023}. 
By superimposing independent sequences of energy levels, 
\citet{Berry_1984} derived a smooth family of distributions spanning Poisson to Wigner-Dyson distributions.  
The combined distribution has energy level density  $\rho = \sum_i \rho_i $ where each component has 
an energy level density $\rho_i$. 
On Figure \ref{fig:2hists}b we plotted (with a dot-dashed blue line) 
a three component model (using equation 33  by 
\citet{Berry_1984})  that includes 
a Poisson component with energy level density $\rho_1=0.1$ and two chaotic components 
with the same energy level density.    We also show with a dotted violet line a Brody distribution 
with power law parameter $\beta = 0.3$.  
The Brody distribution \citep{Brody_1981} is $p_\beta(s)  = d(\beta+1)  s^\beta e^{-b s^{(\beta+1)}}$ with $b  = \left( \Gamma\left(\frac{\beta+2}{\beta+1}\right)\right)^{\beta+1}$
and also lies between Poisson (at $\beta=0$) and Wigner-Dyson (at $\beta=1$) distributions. 
These two curves illustrate that the quasi-energy level spacing distributions of
an ergodic subspace could be modeled with a random matrix model that  
exhibits statistics intermediate between Poisson and Wigner-Dyson distributions.  


In summary, 
Figure  \ref{fig:2hists}a illustrates that the distributions of quasi-energy level spacing in chaotic systems differ from those of regular ones.  Figure \ref{fig:2hists}b illustrates that the distributions of quasi-energy level spacing in chaotic subspaces also differ from regular subspaces.   The spacing distributions are often 
not good matches to either Poisson or Wigner-Dyson distributions, as is typical of mixed systems
 (e.g., \citep{Izrailev_1989,Backer_2011}). 

\section{Harper Hamiltonian in terms of the clock and shift operators \label{sec:clock_shift}}

We write the periodically perturbed Harper Hamiltonian (equation \ref{eqn:Hquantum}) 
in terms of the clock and shift operators $\hat X, \hat Z$, defined in equation \ref{eqn:XZdefs}.
Using our momentum and angle operators $\hat p$, $\hat \phi$, as defined in equations 
\ref{eqn:phihat} and \ref{eqn:phat}, it is convenient to compute 
\begin{align}
\cos \hat p &= \sum_{k=0}^{N-1} \cos \frac{2 \pi k}{N} \ket{k}_F\bra{k}_F \nonumber \\
&= \frac{1}{2} \sum_{k=0}^{N-1} (\omega^k + \omega^{-k}) \ket{k}_F\bra{k}_F = 
\frac{1}{2}(  \hat X + \hat X^\dagger)  \label {eqn:ZZ}\\
\cos \hat \phi &= \frac{1}{2} \sum_{k=0}^{N-1} (\omega^j+ \omega^{-j}) \ket{j} \bra{j} f
= \frac{1}{2} (\hat Z + \hat Z^\dagger) . \label{eqn:XX}
\end{align}

The unperturbed term of equation \ref{eqn:Hquantum} 
can be written in terms of the clock and shift operators 
\begin{align}
\hat h_0 &= a(1- \cos \hat p) - \epsilon \cos \hat \phi \nonumber \\
& = a - \frac{a}{2}(  \hat X + \hat X^\dagger) -  \frac{\epsilon }{2} (\hat Z + \hat Z^\dagger).
\label{eqn:hath}
\end{align} 
In the basis with $\hat \phi$ diagonal, 
$\hat h_0$ is a real, symmetric and periodic tridiagonal matrix
\begin{align}
\hat h_0 = \begin{pmatrix} 
\alpha_0 & \beta & 0 & 0 & 0 & \ldots  & \ldots&0 & \beta \\
\beta &\alpha_1 & \beta & 0 & 0 & \ldots& \ldots&0 & 0 \\
0     & \beta  &\alpha_2 & \beta  & 0  &\ldots &\ldots & 0 & 0 \\
  \vdots    &&  \ddots  &&   \vdots  & &\ddots & &\vdots   \\
0     &  0     & \ldots  & \beta & \alpha_j & \beta & \ldots & 0  & 0 \\
  \vdots    && \ddots  &&    \vdots  && \ddots  && \vdots   \\
0 & 0 & 0& \ldots   & \ldots  & \beta & \alpha_{N-3} & \beta & 0 \\
0 & 0 & 0 & \ldots  & \ldots  & 0 &\beta & \alpha_{N-2} & \beta  \\
\beta & 0& 0& \ldots   & \ldots  &  0 & 0& \beta & \alpha_{N-1}  \\
 \end{pmatrix}
\end{align}
with off-diagonal elements $\beta = -a/2$ and diagonal elements $\alpha_j =a -\epsilon \cos \frac{2 \pi j}{N}$ with $j \in \{0, 1, ... , N-1\}$. 

The perturbation terms 
\begin{align} 
\hat h_1(\tau) & = -\mu \cos (\hat \phi - \tau) - \mu' \cos (\hat \phi + \tau)  \nonumber \\
& = -\frac{(\mu + \mu')}{2} (\hat Z + \hat Z^\dagger) \cos \tau  \nonumber \\
& \qquad -  \frac{(\mu - \mu')}{2i }(\hat Z - \hat Z^\dagger)\sin\tau.  \label{eqn:hath1}
\end{align}

The Harper model has the nice property that it is simply written in terms 
of the shift and clock operators.  In the form of equation \ref{eqn:hath} and in the $N \to \infty$ limit, 
(see \citet{Strohmer_2021}) 
$\hat h_0$ is equivalent to the operator known as the {\it almost Mathieu operator} 
\citep{Simon_1982}. 

The clock and shift operators obey the following relations 
\begin{align}
\hat Z \hat X &= \omega \hat X \hat Z \\
\hat Z^N &= \hat X^N = \hat I \\ 
\hat Q_{FT} \hat X Q_{FT}^\dagger &= \hat Z  
\label{eqn:QXQ} \\
\hat Q_{FT}^\dagger \hat X Q_{FT} &= \hat Z^\dagger 
\end{align}
where $\hat I$ is the identity operator and $\omega = e^{2 \pi i/N}$. 
These identities aid in computing various commutators. 

Since our perturbation depends upon $\cos \hat \phi$, 
and our Hamiltonian contains $\cos \hat p$,  we compute the commutators, 
\begin{align}
[ \cos \hat p,  \cos \hat \phi ] & =  ((\cos \phi_N -1) \cos \hat \phi -  \sin \phi_N \sin \hat \phi ) \cos \hat p
\nonumber \\
[ \cos \hat p,  \sin \hat \phi ]  & = ((\cos\phi_N -1) \sin \hat \phi + \sin \phi_N \cos \hat \phi ) \cos \hat p 
,
\end{align}
where 
\begin{align} 
\phi_N = \frac{2 \pi}{N} . 
\end{align}
In the limit of large $N$
\begin{align}
[ \cos \hat p,  \cos \hat \phi ] & \sim - \frac{2 \pi}{N} \sin \hat \phi  \cos \hat p\nonumber \\
[ \cos \hat p,  \sin \hat \phi ]  & \sim \frac{2\pi}{N} \cos \hat \phi \cos \hat p 
. \label{eqn:coms}
\end{align}

\section{The energy dispersion computed from eigenstates of a perturbed system \label{sec:pert}}

Consider a Hermitian static (time independent) operator $\hat h_0$ that is perturbed with another static 
Hermitian operator $\hat A$  
\begin{align}
\hat h = \hat h_0 + \mu \hat A \label{eqn:pert}
\end{align}
in a finite dimensional Hilbert space and with small real perturbation parameter $\mu$.
We take eigenstates of $\hat h_0$ with energy $E_j$ to be $\ket{v_j}$  (here indexed by integer $j$) 
and those of $\hat h$ to be the set $\ket{w_k}$,  here indexed by integer $k$. 
The indices span 0 to $N-1$ where $N$ is the dimension of the discrete quantum space. 
Conventional non-degenerate quantum perturbation theory  
gives an expression for the eigenstates of $\hat h$ 
as a series expansion in powers of $\mu$. To second order in $\mu$  
\begin{align}
\ket{w_j} & = (1 + a_j) \ket{v_j}  + \sum_{k \ne j} b_{jk} \ket{v_k}+ {\cal O} (\mu^3) 
\end{align}
with \begin{align}
a_j & = - \frac{\mu^2}{2} \sum_{k\ne j} \frac{|A_{kj} |^2  }{E_{jk}^2} \\
b_{jk} & = \mu  \frac{  A_{kj} }{E_{jk} } + \mu^2 \sum_{l \ne j} \frac{ A_{kl} A_{lj}} {E_{jk} E_{jl} } 
 - \mu^2  \frac{  A_{kj} A_{jj} }{ E_{jk}^2} ,
\end{align}
using shorthand $A_{jk} = \bra{v_j} \hat A \ket{v_k}$ and $E_{jk} = E_j - E_k$. 
To second order in $\mu$ 
\begin{align}
|b_{jk}|^2 & = \mu^2  \frac{  |A_{kj} |^2}{E_{jk}^2 }  \\ 
a_j + a_j^* & = - \mu^2 \sum_{k\ne j} \frac{  |A_{kj} |^2}{E_{jk}^2 } . 
\end{align}
We use these expressions to compute the expectation values of the unperturbed operator $\hat h_0$ 
using the eigenstates of $\hat h$, 
\begin{align}
\mu_{h0,j} & = \bra{w_j} \hat h_0 \ket{w_j} \nonumber \\
 & = (1 + a_j + a_j^* )E_j + \sum_{k\ne j} |b_{jk}|^2 E_k \nonumber \\
& = E_j - \mu^2   \sum_{k\ne j} E_{jk}  \frac{  |A_{kj} |^2}{E_{jk}^2 }  \label{eqn:mu} \\
 \mu_{h0,j}^2 
& = E_j^2 - 2 E_j \mu^2 \sum_{k \ne j} \frac{ |A_{kj}|^2}{E_{jk}} 
 \end{align}
 \begin{align}
\bra{w_j} \hat h_0^2  \ket{w_j}
& = (1 + a_j + a_j^* )E_j^2+ \sum_{k\ne j} |b_{jk}|^2 E_k^2 \nonumber \\
& = E_j^2 - \mu^2 \sum_{k\ne j}  (E_j + E_k)  \frac{  |A_{kj} |^2}{E_{jk} }  . 
\end{align}
Putting these together, the dispersion
\begin{align}
\sigma_{h0,j}^2 & = \bra{w_j} \hat h_0^2  \ket{w_j}  - (\bra{w_j} \hat h_0 \ket{w_j})^2 \nonumber \\
 & = \mu^2 \sum_{k \ne j}|A_{jk}|^2 . 
\end{align}
Note that $\mu_{h0,j}$ (equation \ref{eqn:mu}) is second order in $\mu$ but also depends on the inverse of
the energy differences, whereas $\sigma_{h0,j}$ is first order in $\mu$ but only depends on 
the matrix elements. 

To summarize, 
to second order in the perturbation parameter $\mu$, and with no degeneracies, the energy 
standard deviation 
(of the unperturbed Hamiltonian) 
computed with the eigenstates of the perturbed operator $\hat h$ 
\begin{align}
\sigma_{h0,j} 
&  = \mu \sqrt{ \sum_{j \ne k} | \bra{v_j} \hat A \ket{v_k} |^2}. \label{eqn:second}
\end{align}
The term on the right is independent of the energy level differences and depends only
the matrix elements of the perturbation  
$\hat A$ computed with the eigenstates of the unperturbed operator $\hat h_0$.

\end{document}